\documentclass[prl, lettersize,twocolumn,superscriptaddress,
preprintnumbers,nofootinbib,showpacs]{revtex4-1}
\usepackage{graphicx}
\usepackage{amsmath}
\usepackage{amssymb}
\usepackage{color}
\usepackage{hyperref}
\usepackage{url}
\usepackage{breakurl}
\usepackage{float}
\usepackage{epstopdf}
\usepackage{extarrows}
\usepackage[normalem]{ulem}

\newcommand{\beq}{\begin{equation}}
\newcommand{\eeq}{\end{equation}}
\newcommand{\be}{\begin{equation}}
\newcommand{\ee}{\end{equation}}
\newcommand{\bea}{\begin{eqnarray}}
\newcommand{\eea}{\end{eqnarray}}
\newcommand{\nn}{\nonumber}

\begin{document}
\preprint{
{\vbox {
\hbox{\bf MSUHEP-19-009}
}}}
\vspace*{0.2cm}

\title{Resolving the degeneracy in top quark Yukawa coupling with Higgs pair production}

\author{Gang Li}
\email{gangli@phys.ntu.edu.tw}
\affiliation{Department of Physics, National Taiwan University, Taipei, Taiwan 10617}

\author{Ling-Xiao Xu}
\email{lingxiaoxu@pku.edu.cn}
\affiliation{Department of Physics and State Key Laboratory of Nuclear Physics and Technology, Peking University, Beijing 100871, China}

\author{Bin Yan}
\email{yanbin1@msu.edu}
\affiliation{Department of Physics and Astronomy, Michigan State University, East Lansing, MI 48824, USA}

\author{C.-P. Yuan}
\email{yuan@pa.msu.edu}
\affiliation{Department of Physics and Astronomy, Michigan State University, East Lansing, MI 48824, USA}

\begin{abstract}
The top quark Yukawa coupling ($y_t$) can be modified by two dimension-six operators $\mathcal{O}_H$ and $\mathcal{O}_y$ with the corresponding Wilson coefficients $c_H$ and $c_y$, whose individual contribution cannot be distinguished by measuring $y_t$ alone. However, such a degeneracy can be resolved with Higgs boson pair production. In this work we explore the potential of resolving the degeneracy of the unknown Wilson coefficients $c_H$ and $c_y$ at the 14~TeV LHC and the 100~TeV hadron colliders. Combining the information of the single Higgs production, $t\bar{t}h$ associated production and Higgs pair production, the individual contribution of $c_H$ and $c_y$ to $y_t$ can be separated. Regardless of the value of $c_H$, the Higgs pair production can give a strong constraint on $c_y$ at the 100 TeV hadron collider. We further show that it is possible to differentiate various $c_y$ and $c_t$ values predicted in several benchmark models.
\end{abstract}

\maketitle

\noindent{\bf Introduction.~}%
Top quark Yukawa coupling ($y_t$) is the only coupling with the magnitude of order one in the Standard Model (SM). 
As the largest Yukawa coupling, it is important for vacuum stability and cosmology~\cite{Degrassi:2012ry,Bezrukov:2014ina}.
Besides, in many new physics (NP) scenarios~\cite{Martin:1997ns,Contino:2010rs,Bellazzini:2014yua,Panico:2015jxa,Csaki:2016kln}, top quark plays an important role in triggering the electroweak symmetry breaking (EWSB) and is directly connected to new physics beyond the SM. Therefore, it is highly motivated to understand the top quark Yukawa sector better, both theoretically and experimentally. 
The parameter $y_t$ can be measured directly by the $t\bar{t}h$ associated production. Recently, this process is confirmed by both the ATLAS and CMS collaborations with  signal strengths 
$\mu_{t\bar{t}h}=1.32_{-0.26}^{+0.28}$ and $1.26^{+0.31}_{-0.26}$~\cite{Aaboud:2018urx,Sirunyan:2018hoz}, respectively, at the Large Hadron Collider (LHC) with $\sqrt{s}=13~\text{TeV}$.  Besides, $y_t$ can also be measured  in loop-induced single Higgs boson production~\cite{ATLAS:2018doi,CMS:2018lkl}, $t(\bar{t})hj$ associated production~\cite{Sirunyan:2018lzm} and multi-top production processes~\cite{Cao:2016wib,Cao:2019ygh}. With higher luminosity being accumulated, one expects the accuracy on $y_t$ can be further improved. It is thus timely to study what kind of NP can modify $y_t$.

In general, we can parameterize NP effects on $y_t$ by several higher dimensional operators in a model independent way. Out of the complete set of dimension-6 operators listed in Ref.~\cite{Pomarol:2013zra}, we consider in this work the two operators which can modify $y_t$ at tree level: 
\bea
\mathcal{L}=\mathcal{L}_{\text{SM}}+c_H \mathcal{O}_H+(c_y \mathcal{O}_y+\text{h.c.})+\cdots
\eea
in the so called Strongly-Interacting Light Higgs (SILH) basis~\cite{Giudice:2007fh,Contino:2013kra}, where the dimension-six operators 
\begin{align}
\mathcal{O}_H&=\dfrac{1}{2v^2}\partial^\mu(H^{\dagger}H)\partial_\mu(H^{\dagger}H)\ ,\nn\\
\mathcal{O}_y&=-\dfrac{y_t^{\rm SM}}{v^2}H^{\dagger}H\bar{Q}_L\tilde{H}t_R\ .
\label{operator}
\end{align}
Here, $c_H$ and $c_y$ are the corresponding Wilson coefficients with $c_y$ being assumed to be real, $Q_L$ is the left-handed third-family quark doublet, $v=246~{\rm GeV}$ is the vacuum expectation value, $y_t^{\rm SM}=m_t/v$ is the SM top Yukawa coupling, and $m_t$ is the top quark mass. As $\mathcal{O}_H$ modifies the Higgs boson wave function, it can universally shift all the single Higgs couplings, hence, affects $y_t$.

Theoretically, the operators $\mathcal{O}_H$ and $\mathcal{O}_y$ can be induced by several different NP scenarios. 
For example, scalar singlets interacting with the Higgs doublet can induce the universal $\mathcal{O}_H$ operator~\cite{Craig:2013xia,Craig:2014una,Dawson:2017vgm,Corbett:2017ieo,Cao:2017oez}, while additional vector-like fermions can induce the operator $\mathcal{O}_y$ via mixing with the top quark~\cite{delAguila:2000rc,Chen:2017hak}. 
As both the operators in Eq.~\eqref{operator} can induce deviations in $y_t$, one cannot differentiate their individual contributions if we only measure the top Yukawa coupling. 
Even if $y_t$ is measured to be consistent with the SM prediction,
one cannot exclude the possibility of having cancellation among different NP operators. Or, if the deviation in $y_t$ is established, we still need to separate the effects of $\mathcal{O}_H$ and $\mathcal{O}_y$ for better understanding the origin of NP.
Since the effect of the $\mathcal{O}_H$ operator is to simply rescale any amplitude involving a single Higgs boson by a factor $1/\sqrt{1+c_H}$ after renormalizing the Higgs boson field, its effect is universal and can be measured from studying the  
$hVV~(V=W^\pm,Z)$ couplings. However as shown in Ref.~\cite{Cao:2018cms}, in case that Higgs boson is a pseudo Nambu-Goldstone boson, other operator (at the order of $\mathcal{O}(p^2)$) can mimic the effect induced by $\mathcal{O}_H$ on $hVV$ couplings. Hence, it requires novel method to separately measure the coefficients of those two types of operators~\cite{Cao:2018cms}. 
Likewise, in this work, we explore the possibility of separately measuring the coefficients of $\mathcal{O}_H$ and $\mathcal{O}_y$, both contributing to top Yukawa coupling $y_t$, through Higgs boson pair production $gg\to hh$. 
In addition to modifying the single Higgs effective coupling of $ht\bar{t}$, both $\mathcal{O}_H$ and $\mathcal{O}_y$ can also contribute to the effective coupling $hht\bar{t}$, but with different combinations, which can be measured via Higgs boson pair production. 
Namely, studying Higgs boson pair production can be utilized not only for measuring Higgs self interactions, but also for discriminating new physics scenarios in the top sector.
Only after the individual contribution of each effective operator is extracted can we further solidify the SM or otherwise establish NP models.

\noindent{\bf Higgs boson pair production.~}%
The gluon-initiated Higgs pair production can be used to measure the trilinear Higgs self-coupling
\cite{Glover:1987nx, Baur:2003gpa,Baur:2003gp,Dolan:2012rv,
Baglio:2012np,Papaefstathiou:2012qe,Shao:2013bz,Goertz:2013kp,
Barger:2013jfa,Barr:2013tda,deLima:2014dta,Li:2015yia,Behr:2015oqq,
Alves:2017ued,Adhikary:2017jtu,
Kim:2018uty,Goncalves:2018yva,Chang:2018uwu,Borowka:2018pxx,
Homiller:2018dgu,Kim:2018cxf,Kim:2019wns}.
It is also sensitive to various NP models~\cite{Agashe:2004rs,Contino:2006qr,ArkaniHamed:2001nc,
ArkaniHamed:2002qy,Dawson:2017vgm,Corbett:2017ieo,Huang:2017nnw,
Basler:2018dac,Babu:2018uik,Li:2019ghf}.
After the EWSB, the effective Lagrangian related to 
the non-resonant Higgs pair production is~\cite{Cao:2015oaa,Cao:2016zob,Chen:2014xra,
Goertz:2014qta,Azatov:2015oxa,Dawson:2015oha}
\bea
\label{eq:effective}
\mathcal{L}_{h}&=&-\frac{m_h^2}{2v}c_3h^3
-\frac{m_t}{v}c_t\bar{t}t h
-\frac{m_t}{v^2}c_{2t}\bar{t}t h^2 \nn\\
&+&\frac{\alpha_s c_g}{12 \pi v}hG_{\mu\nu}^aG^{\mu\nu}_a
+\frac{\alpha_s c_g}{24 \pi v^2}h^2G_{\mu\nu}^aG^{\mu\nu}_a~,
\label{eq:eff}
\eea
where $a$ is the color index, $\alpha_s=g_s^2/4\pi$ with $g_s$ being the strong coupling strength, and $m_h$ is the Higgs boson mass. 
The SM, at tree level, corresponds to $c_3=c_t=1$ and $c_{2t}=c_g=0$. 
Then the squared amplitude of $gg\to hh$, after averaging over the gluon polarizations and colors, is~\cite{Dawson:2015oha}
\bea
\overline{|{\cal M}|^2} = \frac{\alpha_s^2 \hat{s}^2}{256\pi^2 v^4}
\bigg[ && \bigg|  {3m_h^2\over {\hat s}-m_h^2} c_3
\left(c_t F_\bigtriangleup+{2\over 3} c_g\right) 
 +2c_{2t} F_\bigtriangleup \nn\\ 
&&+c_t^2 F_\Box+{2\over 3}c_g \bigg|^2+ \left|c_t^2  G_\Box\right|^2\bigg],
 \label{eq:hhdif}
\eea
where $F_\bigtriangleup\equiv F_\bigtriangleup({\hat s},{\hat t},m_h^2, m_t^2)$, 
$F_\Box\equiv F_\Box({\hat s}, {\hat t}, m_h^2, m_t^2)$ and 
$G_\Box\equiv G_\Box({\hat s},{\hat t}, m_h^2, m_t^2)$ are the 
form factors~\cite{Plehn:1996wb} with $\hat{s}$ and 
$\hat{t}$ being the canonical Mandelstam 
variables.
The first term inside the bracket contributes to $s$-wave and the $G_\Box$ term to $d$-wave component 
whose contribution to total cross section is numerically negligible~\cite{Dawson:2012mk}.

To avoid any momentum dependent contributions to the Higgs self couplings induced by $\mathcal{O}_H$, we adopt the generalized canonical normalization of the Higgs filed 
and perform the field redefinition~\cite{Plehn:2009nd,Goertz:2014qta}:
\bea
h\to \frac{h}{\sqrt{1+c_H}}-\dfrac{c_H h^2}{2(1+c_H)^2v},\quad
m_t\to \frac{m_t}{1+\frac{1}{2}c_y}.
\eea 
Applying this shift of the Higgs field throughout the Standard Model Lagrangian $\mathcal{L}_{\text{SM}}$ leads to multiple Higgs couplings to any pair of massive gauge bosons or massive fermions.
The effective couplings $c_{t}$ and $c_{2t}$ are derived as
\bea
c_{t}=\dfrac{2+3c_y}{\sqrt{1+c_H}(2+c_y)},\quad 
c_{2t}=\dfrac{c_H(3c_y-2)+6c_y}{2(c_H+1)^2(c_y+2)}.
\eea
We note that $c_t$ and $c_{2t}$ have different dependence on the Wilson coefficients $c_H$ and $c_y$. 
Also, $c_t=1-c_H/2+c_y$ and 
$c_{2t}=3c_y/2-c_H/2$, when $c_y,c_H\ll 1$~\cite{Azatov:2015oxa,Goertz:2014qta}.

\begin{figure}
\includegraphics[scale=0.35]{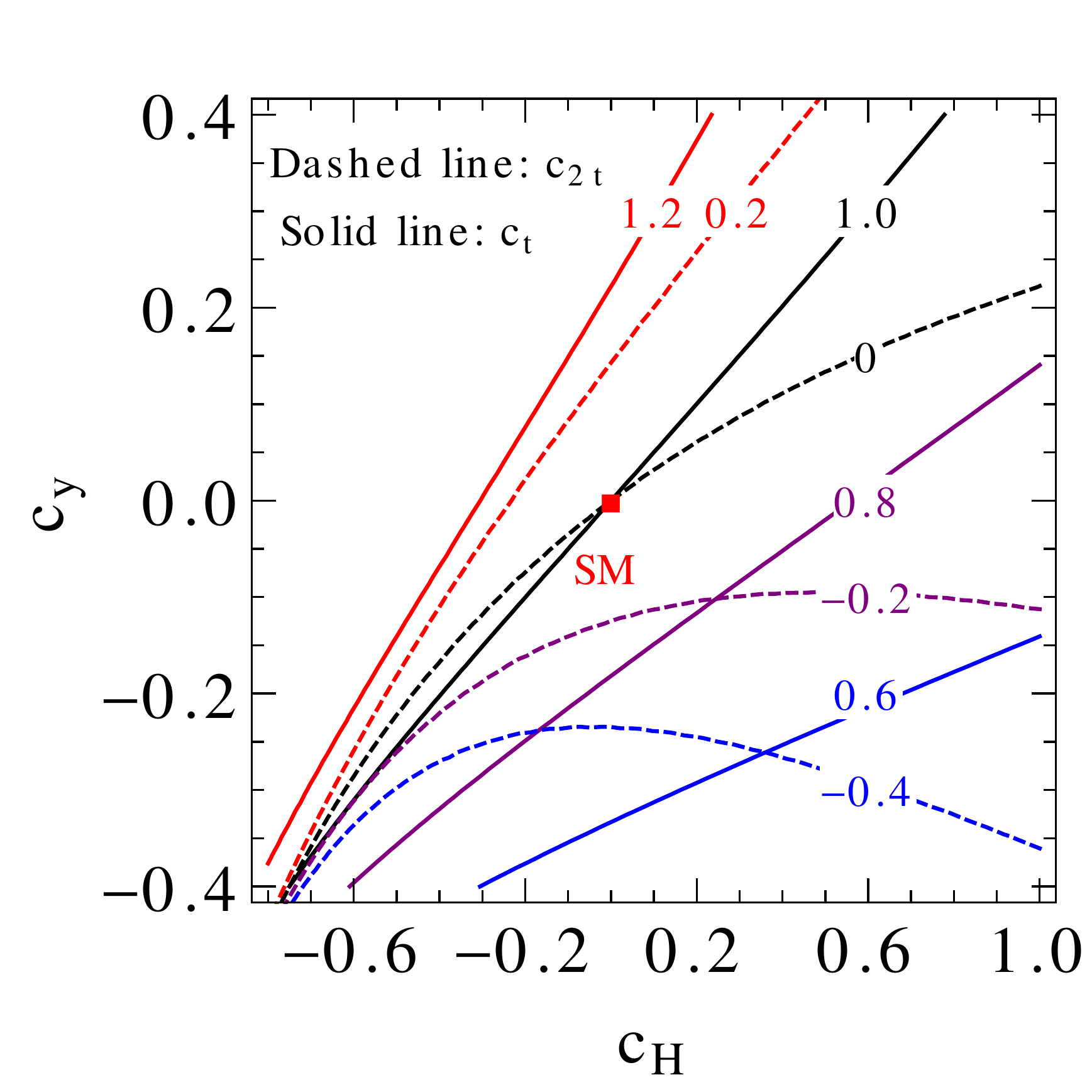}     
\caption{The contours of $c_t$ (solid lines) and $c_{2t}$ (dashed lines) in the plane of  $c_H$ and $c_y$. The number on each curve denotes the specific value of the Higgs effective coupling $c_t$ or $c_{2t}$.
\label{fig:ctc2h} }
\end{figure}
In Fig.~\ref{fig:ctc2h} the dependence of $c_t$ and $c_{2t}$ on $c_H$ and $c_y$ is shown. It is clear that the slope of $c_{2t}$ (dashed lines) is different from $c_t$ (solid lines), especially when $c_{2t}<0$ and $c_t<1$. Precise study on the effective coupling $c_{2t}$ therefore offers the possibility to discriminate the effects of $c_H$ and $c_y$. 

\begin{figure}
\includegraphics[scale=0.35]{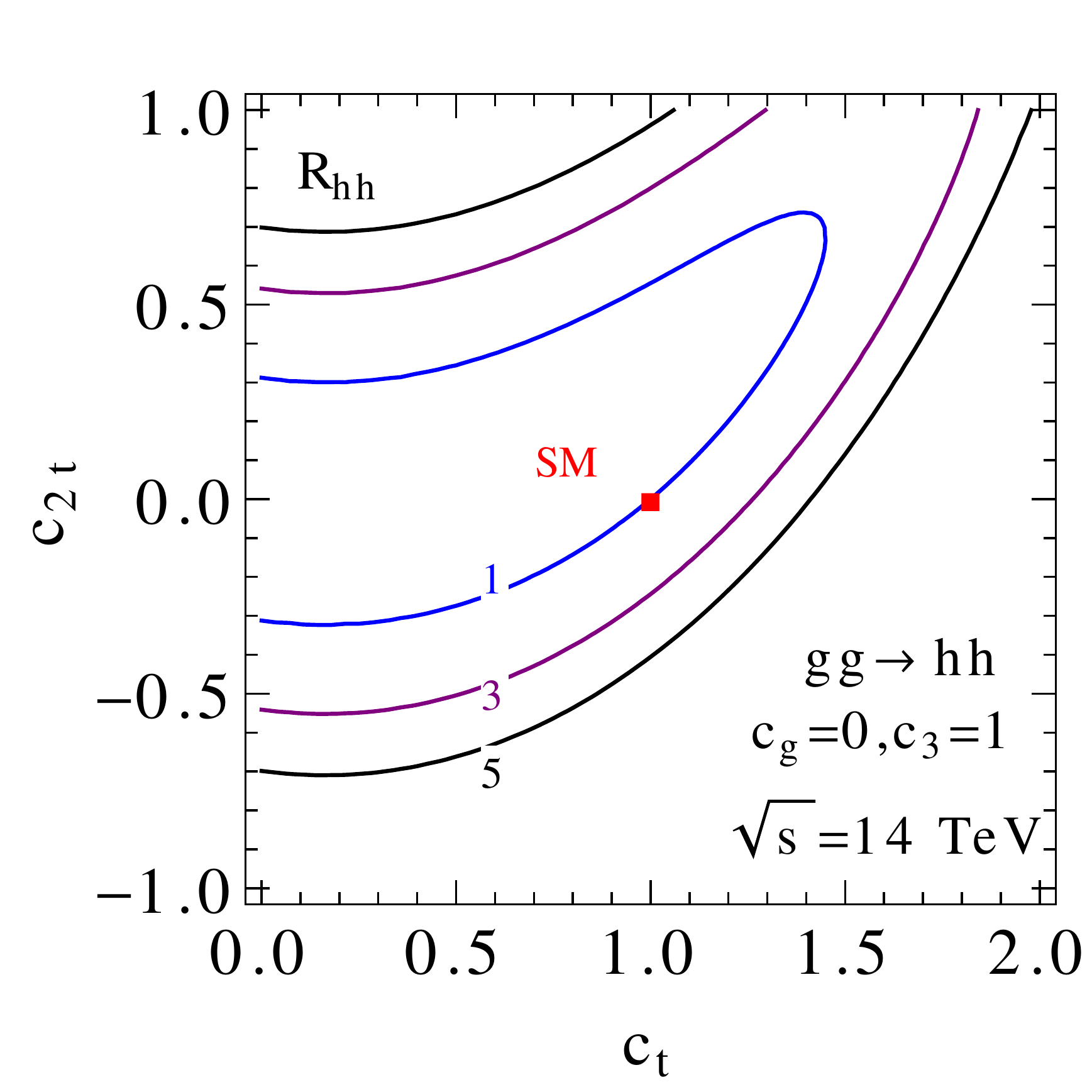}     
\caption{The contours of $R_{hh}$ in the plane of  $c_t$ and $c_{2t}$, with $c_g=0$ and $c_3=1$, at the 14 TeV LHC. 
The red box denotes the SM values ($c_t=1$, $c_{2t}=0$).
\label{fig:ctc2ha} }
\end{figure}

\begin{figure}
\includegraphics[scale=0.24]{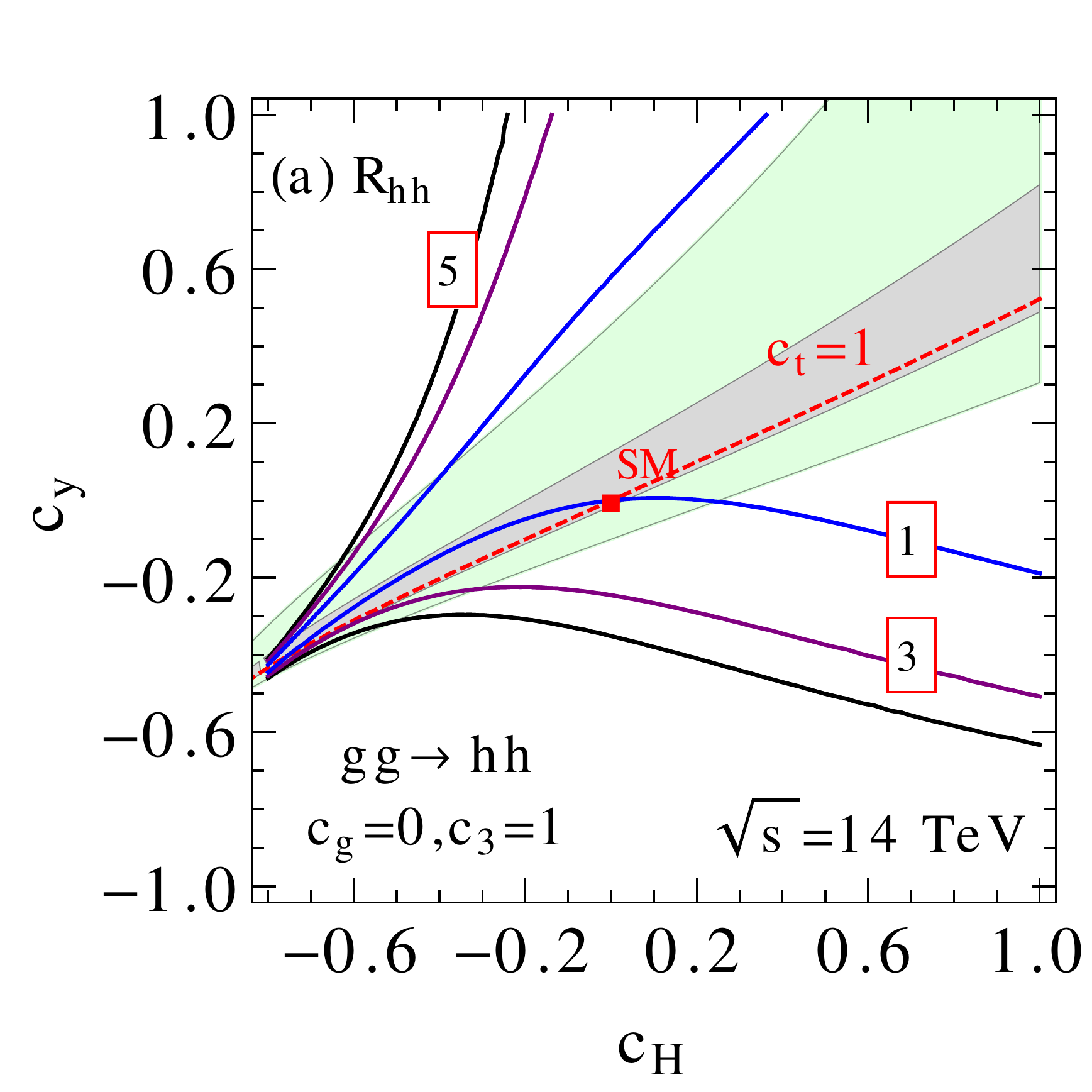}  
\includegraphics[scale=0.24]{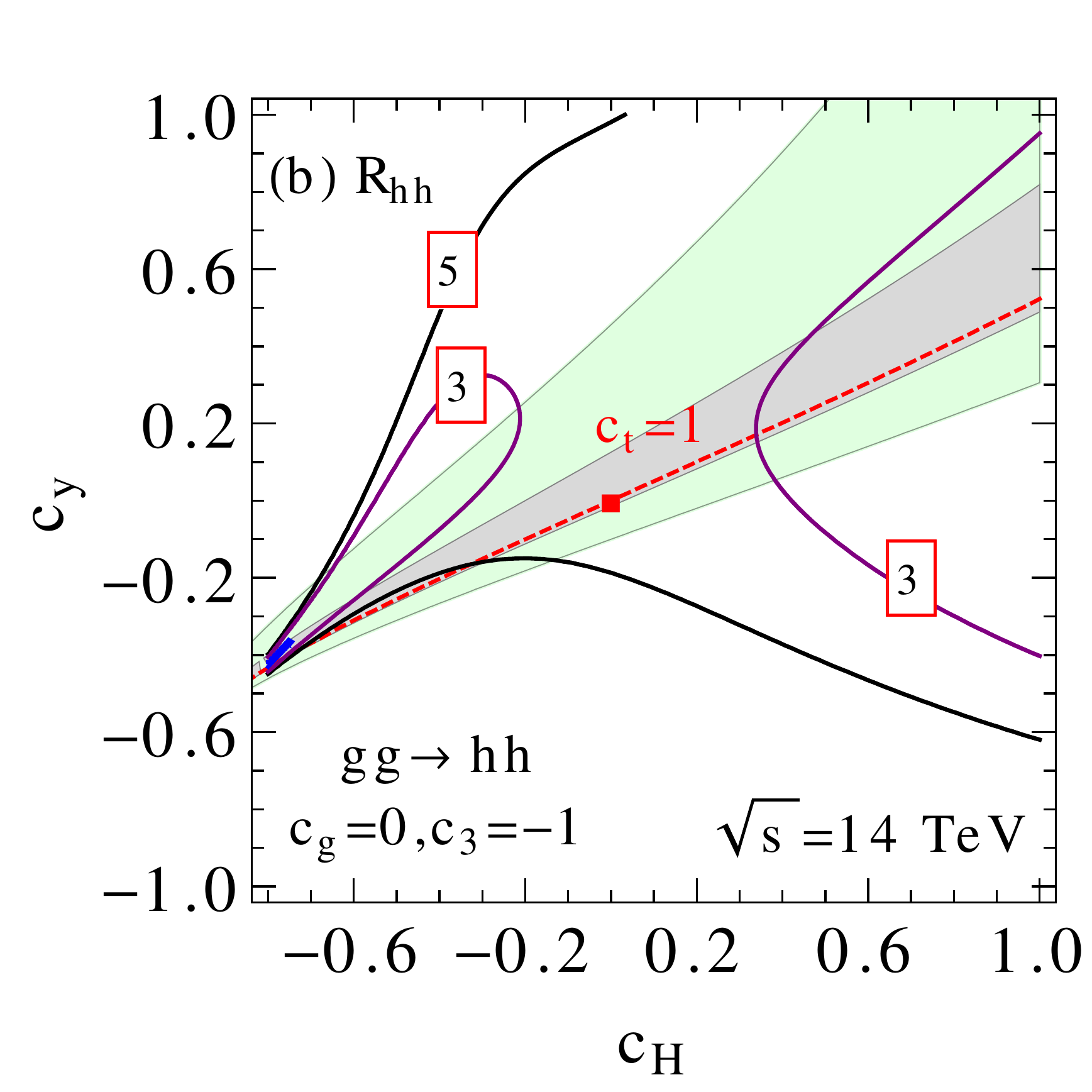}  
\includegraphics[scale=0.24]{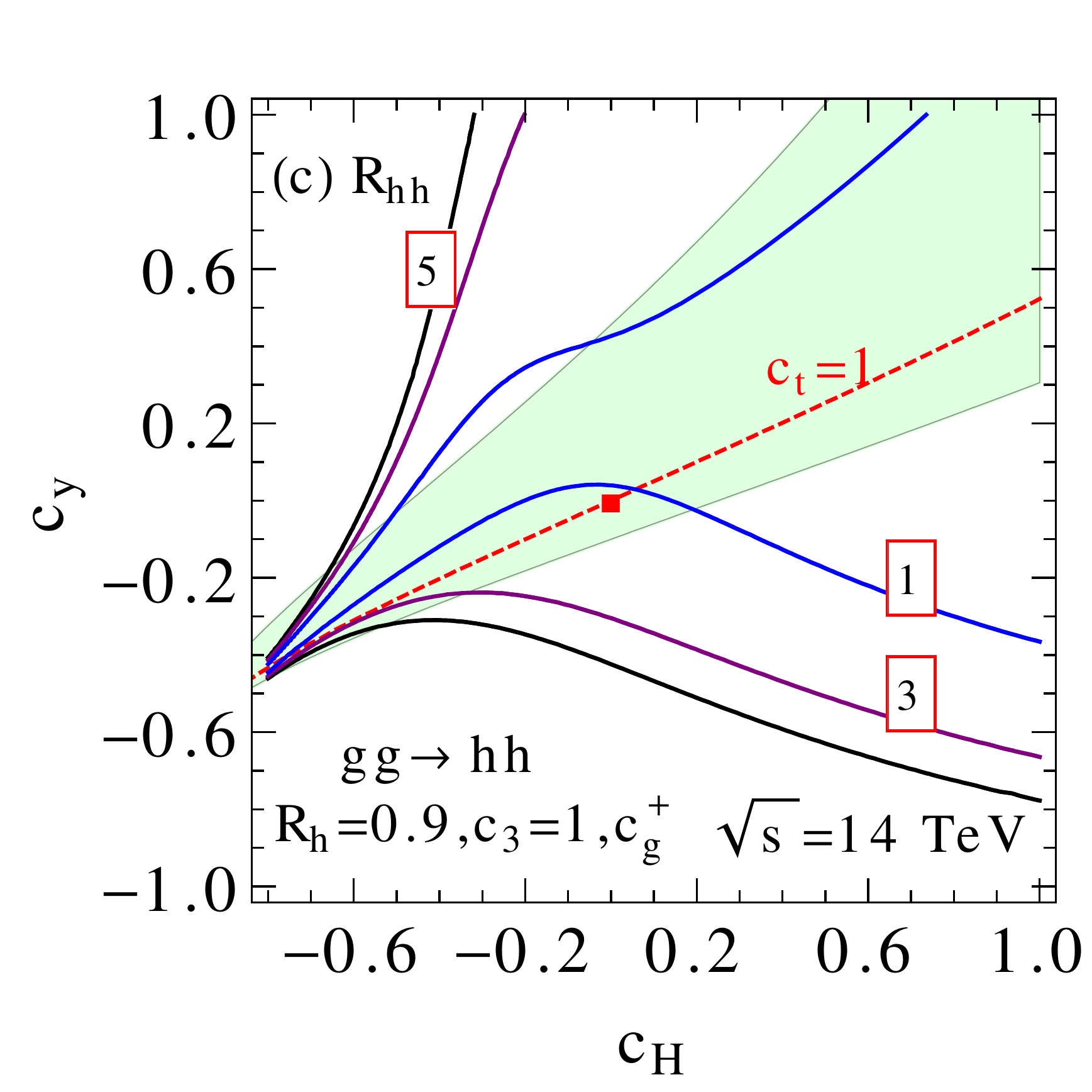}
\includegraphics[scale=0.24]{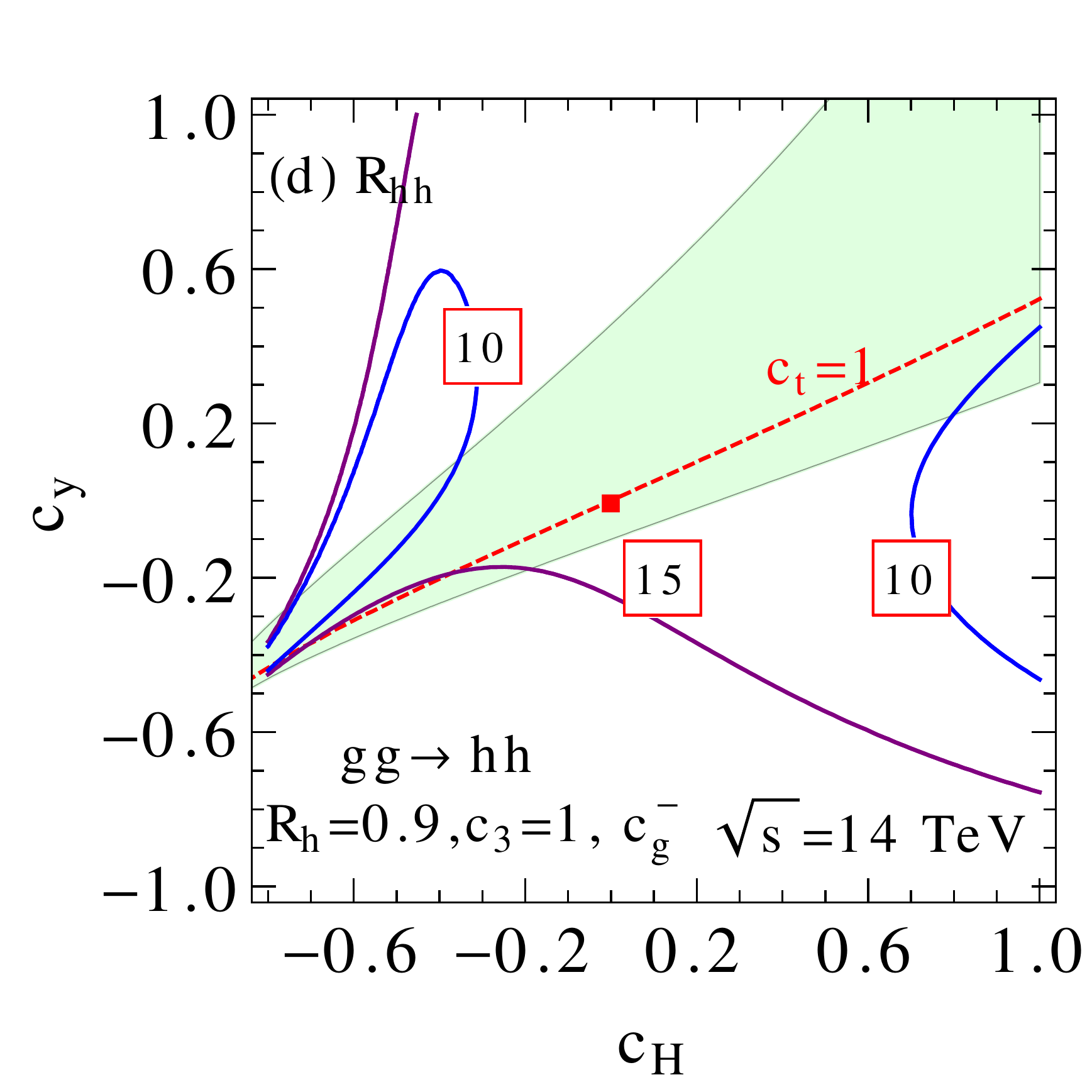}
\includegraphics[scale=0.24]{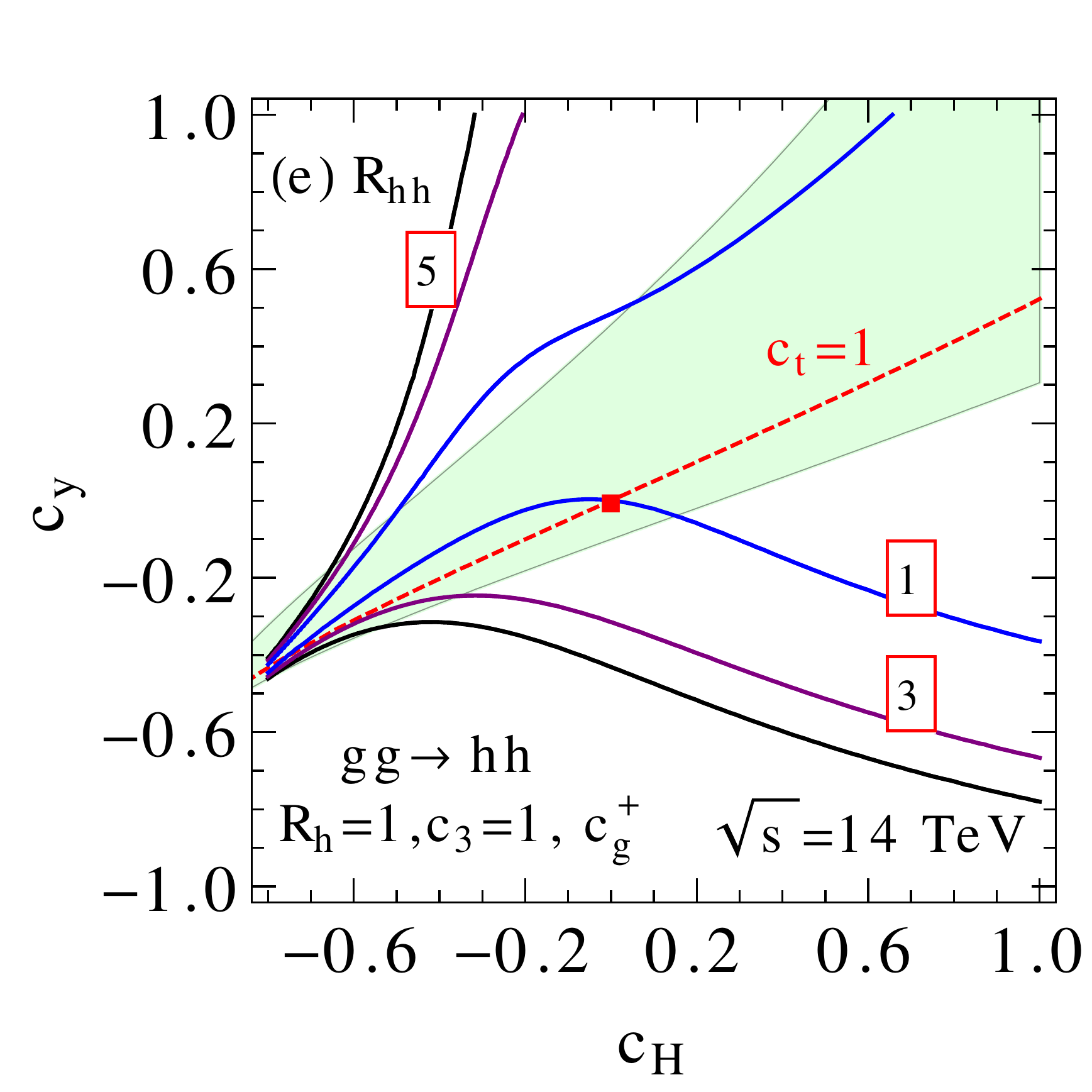}   
\includegraphics[scale=0.24]{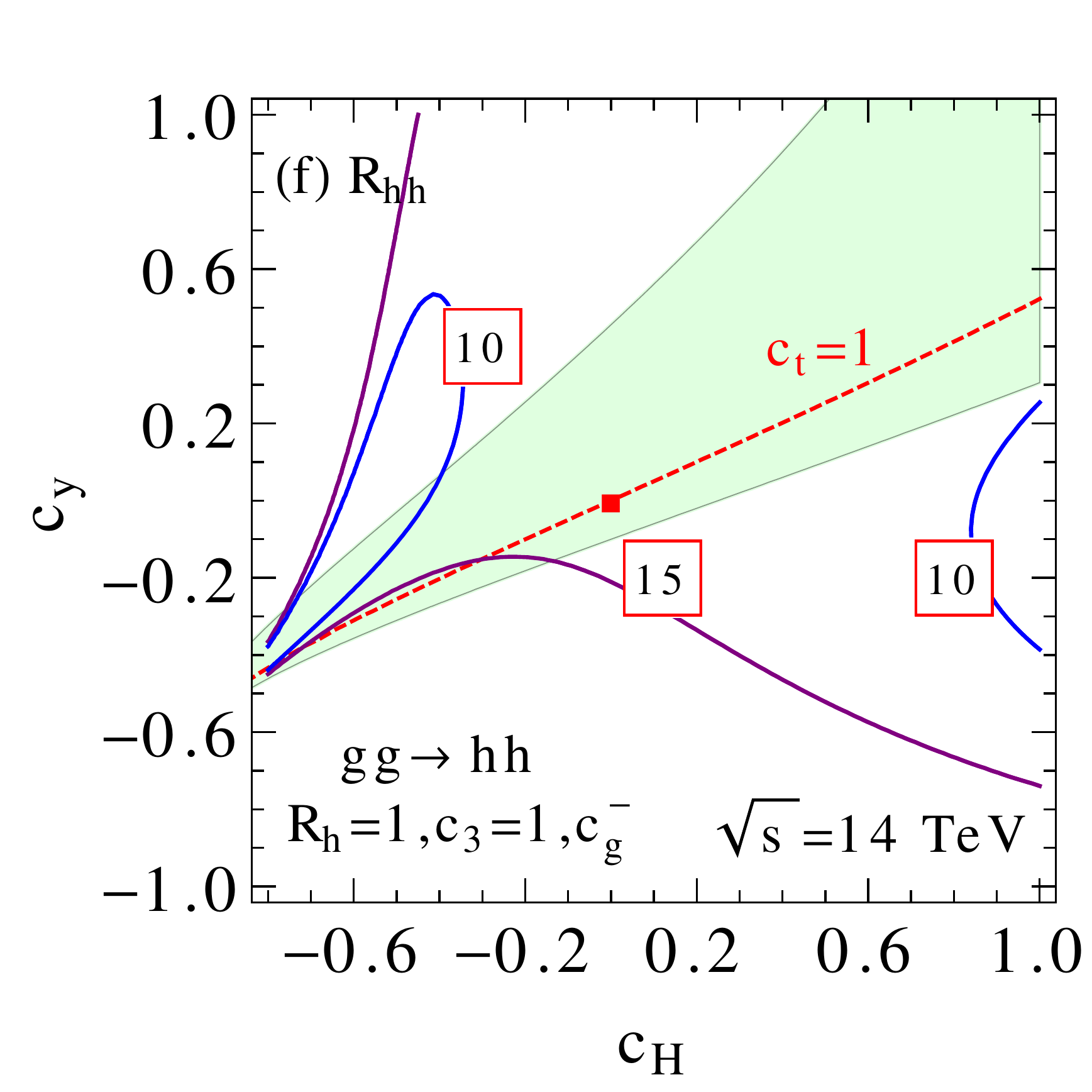}
\includegraphics[scale=0.24]{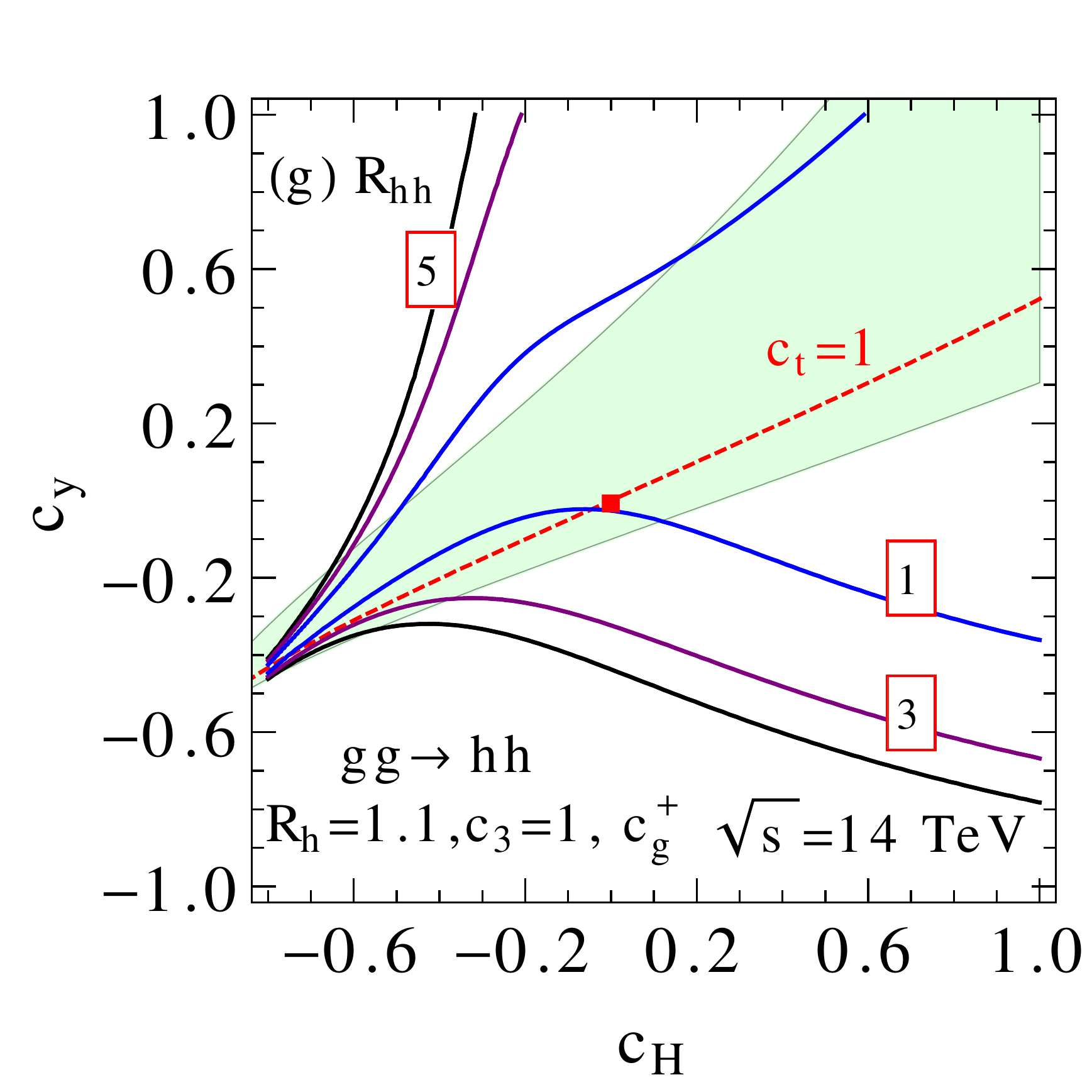}
\includegraphics[scale=0.24]{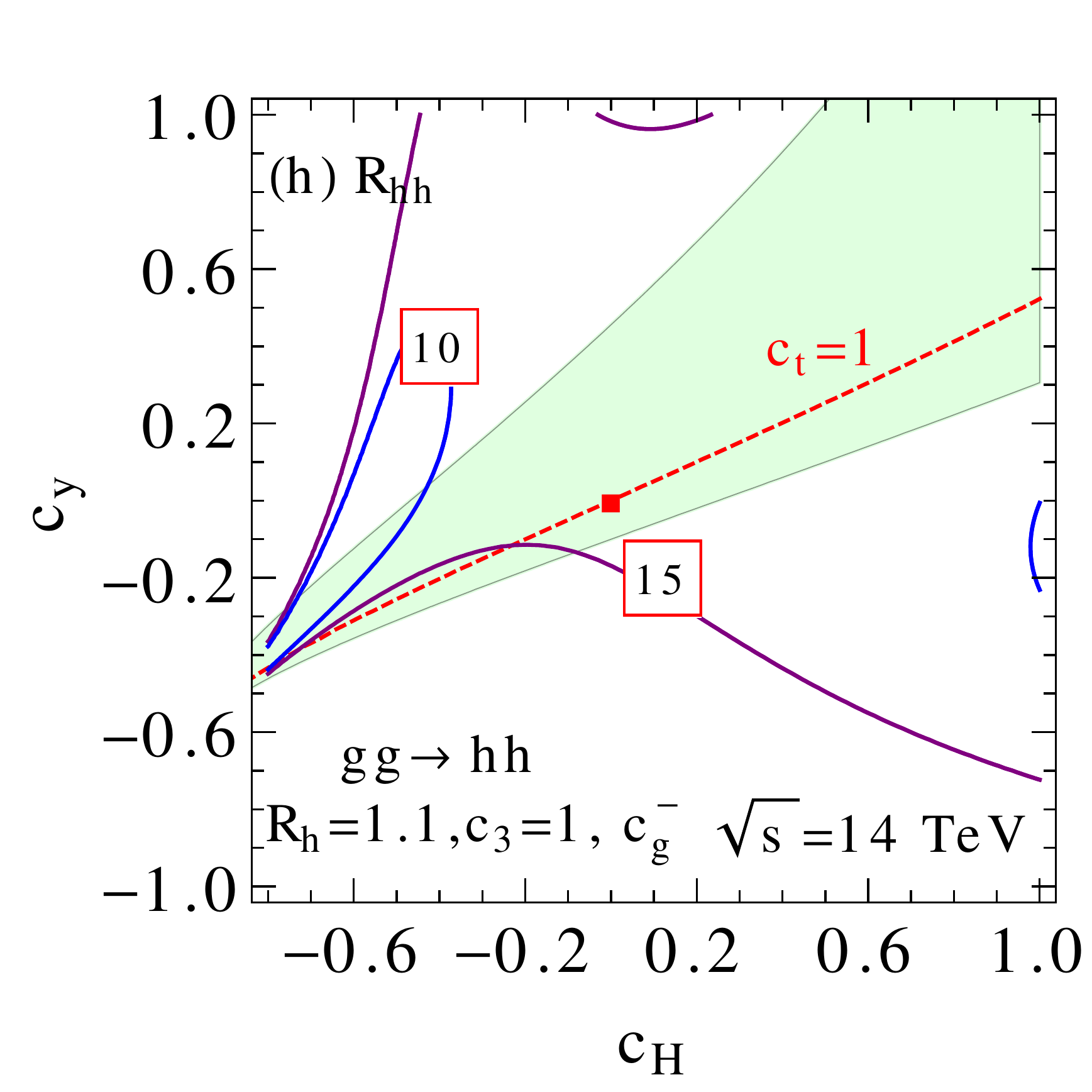}       
\caption{The contours of $R_{hh}$  in the plane of  $c_H$ and $c_y$ at the 14 TeV LHC. The green and gray bands correspond to the constraints, at the $2\sigma$ C.L., from the measurements of $t\bar{t}h$ and single Higgs production cross section, respectively, at the 13 TeV LHC. The red box denotes ($c_H=0$, $c_y=0$), while the red dashed line denotes $c_t=1$. $c_g^{\pm}$ refers to the sign choices "$\pm$" in Eq.~\eqref{eq:cg} with $R_h$ being fit from single Higgs production and decay data.
\label{fig:R14} }
\end{figure}

As shown in Eqs.~\eqref{eq:effective} and \eqref{eq:hhdif}, $c_3$ and $c_g$ also contribute to the Higgs pair production cross section. 
However, $c_g$ is already constrained to be within $c_g^-$ and $c_g^+$ by the signal strength measurements of single Higgs production $gg\to h$~\cite{ATLAS:2018doi,CMS:2018lkl}
\bea
c_g^{\pm}=\frac{3}{2}\left(-c_t F_\bigtriangleup \pm \sqrt{R_h}|F_\bigtriangleup|\right),
\label{eq:cg}
\eea
where $R_h\equiv \sigma(gg\to h)/\sigma^{\rm SM}(gg\to h)$, the sign "$\pm$"  refers to the cases $c_t F_\bigtriangleup>-2/3c_g$ and $c_t F_\bigtriangleup<-2/3c_g$, respectively.  
The combined fit to the single Higgs production, which depends on both $c_t$ and $c_g$~\cite{Cao:2015oaa}, and decay using 13~TeV LHC data gives rise to $R_{h}=1.07\pm 0.09$ (ATLAS~\cite{ATLAS:2018doi}) and $R_h=1.23 \pm 0.13$ (CMS~\cite{CMS:2018lkl}). By fixing $R_h$, we can replace $c_g$ with a function of $c_t$.
Furthermore, the coupling $c_3$ in Eq.~\eqref{eq:eff} is approximately equal to $1-{3 \over 2} c_H + c_6$, as shown in  Refs.~\cite{Azatov:2015oxa,Goertz:2014qta}, and 
is weakly constrained, $-5.0<c_3<12.1$, by present data~\cite{ATLAS:2018otd,Sirunyan:2018two}.
Here, $c_6$ is the coefficient of the dimension-6 operator ${\cal O}_6=\lambda/v^2 |H|^6$.
As the Higgs pair production cross section is more sensitive to the sign of $c_3$, rather than its magnitude~\cite{Cao:2015oxx}, we will choose $c_3=\pm 1$ as our benchmark value in the following numerical analysis. 

To compare $\sigma(gg\to hh)$ with the SM prediction, we define a ratio $R_{hh}$ as
\bea
R_{hh}&=&\frac{\sigma (gg\to hh)}{\sigma^{\rm SM} (gg\to hh)}.
\eea
In Fig.~\ref{fig:ctc2ha}, we show the contours of $R_{hh}$ at the 14 TeV LHC in the plane of $c_t$ and $c_{2t}$, with other parameters chosen as $c_g=0$ and $c_3=1$. $R_{hh}$ can be enhanced largely for both the positive and negative $c_{2t}$, but $R_{hh}$ is more sensitive to negative $c_{2t}$ when $c_t$ is of order one, which can be understood with Eq.~\eqref{eq:hhdif}. In the large top quark mass limit, $F_{\bigtriangleup}\to 2/3$ and $F_{\Box}\to -2/3$~\cite{Plehn:1996wb}. Besides, the $c_t^2 F_{\Box}$ term dominates over the $c_t F_{\bigtriangleup}$ term for $c_{t}\gtrsim 1$. Therefore, a negative $c_{2t}$ can enhance $R_{hh}$ more easily than a positive $c_{2t}$, for this choice of $c_g$ and $c_3$~\cite{Cao:2016zob}.

With the information from Figs.~\ref{fig:ctc2h} and ~\ref{fig:ctc2ha}, we conclude that it is hopeful to discriminate the effects of $c_H$ and $c_y$ through Higgs pair production, especially for the negative $c_{2t}$ region.
Furthermore, we could translate the above results in the plane of $(c_H,c_y)$.
In Fig.~\ref{fig:R14}, we show the contours of $R_{hh}$ with respect to the Wilson coefficients $c_H$ and $c_y$, where various choices of the effective couplings $c_g$ and $c_3$ are considered. They correspond to 
\begin{align}
&(a)\ c_g=0, c_3=1; &  &(b)\ c_g=0, c_3=-1;\nn\\
&(c) \ R_h=0.9, c_3=1, c_g^+; &  &(d)\ R_h=0.9, c_3=1, c_g^-;\nn\\
&(e)\ R_h=1, c_3=1, c_g^+;& & (f)\ R_h=1,c_3=1,c_g^-;\nn\\
&(g)\ R_h=1.1, c_3=1,c_g^+; & &(h)\ R_h=1.1, c_3=1,c_g^-.
\end{align}
For the cases $(a)$ and $(b)$, the $hgg$ effective coupling $c_g$ is assumed to vanish while the trilinear Higgs self-coupling $c_3$ is assumed to be $+1$ and $-1$, respectively.
In these cases, we include the constraints on the parameter space of $c_H$ and $c_y$ from the single Higgs production~\cite{ATLAS:2018doi,CMS:2018lkl}; cf. the gray band of Fig.~\ref{fig:R14} (a, b). It amounts to $c_t=1.03\pm 0.04$ (ATLAS) and $c_t=1.11\pm 0.06$ (CMS).
For the cases from $(c)$ to $(h)$, $c_g$ is derived from a given $R_h$ value, cf. Eq.~\eqref{eq:cg}, while $c_3$ is fixed to be identical to the SM value.
For comparison, we also show the parameter space constrained by the $t\bar{t}h$ measurements at the 13 TeV LHC, with an integrated luminosity of $79.8~{\rm fb}^{-1}$ from ATLAS~\cite{Aaboud:2018urx} and of  $35.9~{\rm fb}^{-1}$ from CMS~\cite{Sirunyan:2018hoz}.
Here, we take the combined best fit $t\bar{t}h$ signal strength normalized to the SM prediction, such that  
$c_t=1.15^{+0.12}_{-0.11}$ (ATLAS) and $c_t=1.12^{+0.14}_{-0.12}$ (CMS).

With the result depicted in Fig.~\ref{fig:R14}, several comments are in order:
\begin{itemize}
\item  $R_{hh}$ is enhanced in some parameter space of $c_H$ and $c_y$, the magnitude of the enhancement strongly depends on the choice of $c_3$ and $c_g$.
\item In case $(b)$, the cancellation between the triangle diagram and box diagram does not happen, because of the negative $c_3$ which further enhances $R_{hh}$~\cite{Baglio:2012np}.
\item In case of $c_g^+$ or $c_g^-$, the value of $c_g$ is extracted from the signal strength of single Higgs production process $R_h$. We find that the variation of $R_h=0.9,\ 1,$ or $1.1$ can only slightly change the value of $R_{hh}$. 
\item  $R_{hh}$ is more sensitive to $c_g^-$ than $c_g^+$. According to Eq.~\eqref{eq:cg}, $c_g^+$ is close to zero for a positive $c_t$, and accordingly the contours of $R_{hh}$ are similar to the contours in case $(a)$. On the other hand, negative $c_g^-$ can significantly deviate from zero and $R_{hh}$ is largely enhanced (cf. Eq.~\eqref{eq:hhdif}).
\item In cases $(a)$, $(c)$, $(e)$ and $(g)$, the possible enhancement of $R_{hh}$ can only come from the deviations of $c_t$ and $c_{2t}$. With the information of Figs.~\ref{fig:ctc2h} and~\ref{fig:ctc2ha}, $R_{hh}$ can be largely enhanced when $c_{2t}<0$, which corresponds to the region $c_y<0$ in Fig.~\ref{fig:R14}.
\end{itemize}

\noindent{\bf Sensitivity at the 14 TeV LHC and the 100 TeV hadron collider.~}%
Now we discuss the potential of discriminating the Wilson coefficients $c_H$ and $c_y$ at the 14 TeV LHC and the 100 TeV proton-proton hadron collider.
As a concrete example, we examine the $b\bar{b}\gamma\gamma$ channel, which has been studied by the ATLAS collaboration at the High-Luminosity LHC (HL-LHC), operating at the center-of-mass energy of $14~ {\rm TeV}$ with an integrated luminosity of $3~{\rm ab}^{-1}$~\cite{ATL-PHYS-PUB-2014-019}.
As being discussed in Refs.~\cite{Cao:2015oaa,Cao:2016zob}, an analytical function can be used to describe the fraction of signal events passing through the kinematic cuts. Since the Higgs boson is a scalar particle and the $gg\to hh$ process is dominated by the $s$-wave contribution, the acceptance of the kinematic cuts,  
in the inclusive Higgs pair production,
will mainly depend on the invariant mass of the Higgs boson pair ($m_{hh}$).
The cross section of $gg\to hh$, after imposing the  kinemati cuts, can be written as~\cite{Cao:2015oaa}
\bea
\sigma_{\text{cut}}=\int dm_{hh}\frac{d\sigma}{dm_{hh}}{\mathcal{A}}(m_{hh})
\label{eq:convolution}
\eea
where the efficiency function $\mathcal{A}(m_{hh})$ has been given in Refs.~\cite{Cao:2015oaa,Cao:2016zob}, both for the 14 TeV LHC and the 100 TeV hadron collider. In Ref.~\cite{Cao:2016zob}, it is demonstrated that the results obtained from the analytic cut efficiency functions agree very well with other results with more detailed simulations~\cite{Azatov:2015oxa}.

\begin{figure}
\includegraphics[scale=0.24]{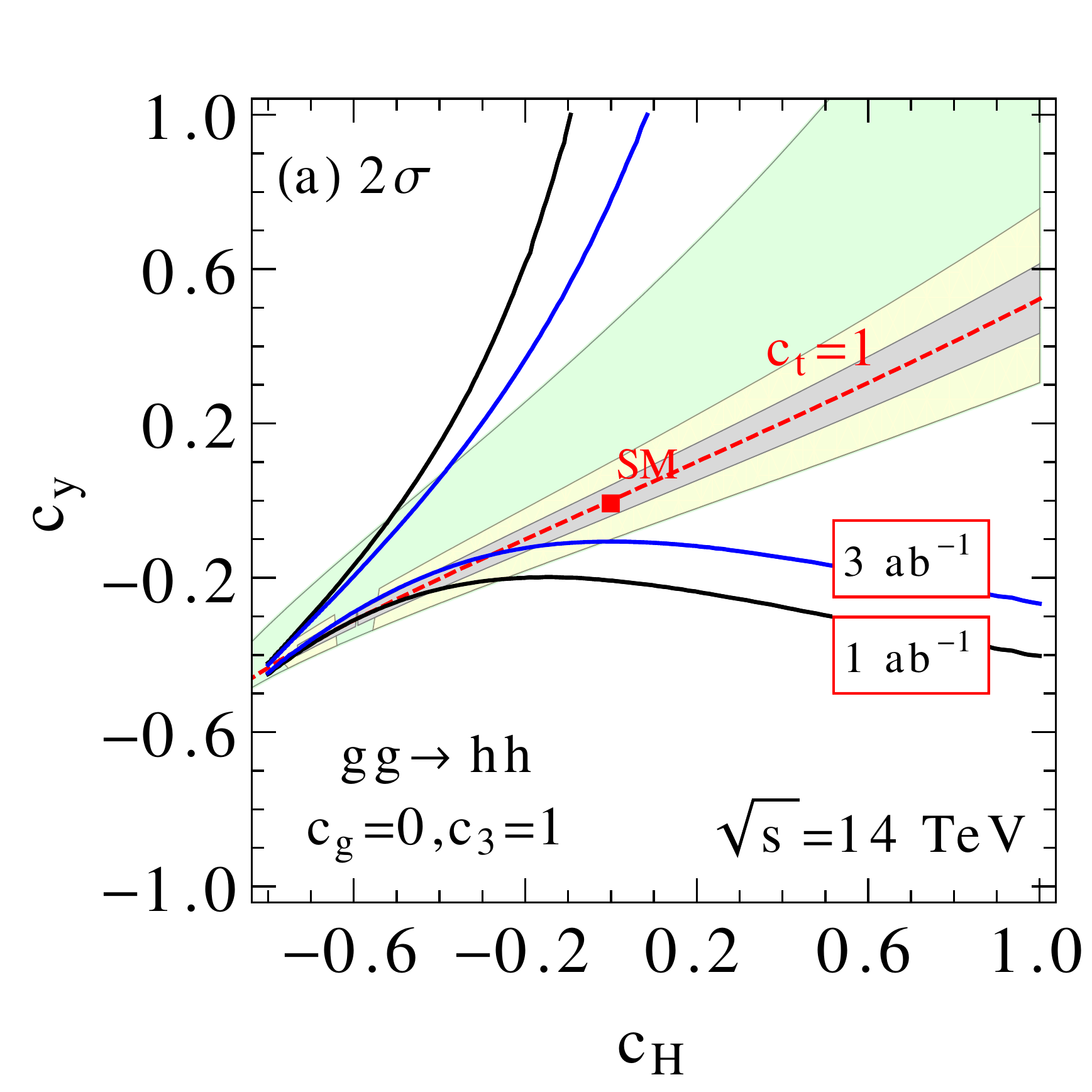}  
\includegraphics[scale=0.24]{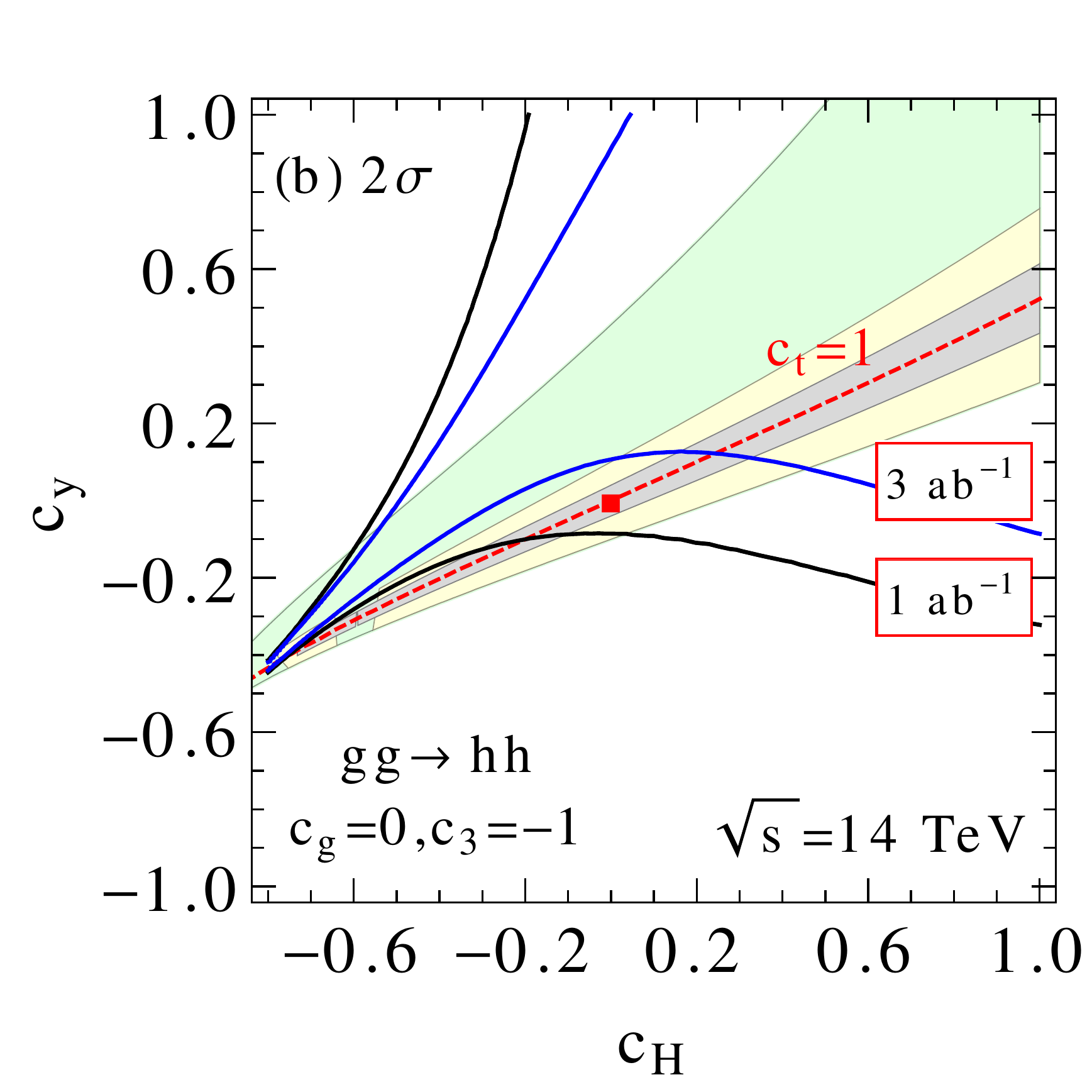}  
\includegraphics[scale=0.24]{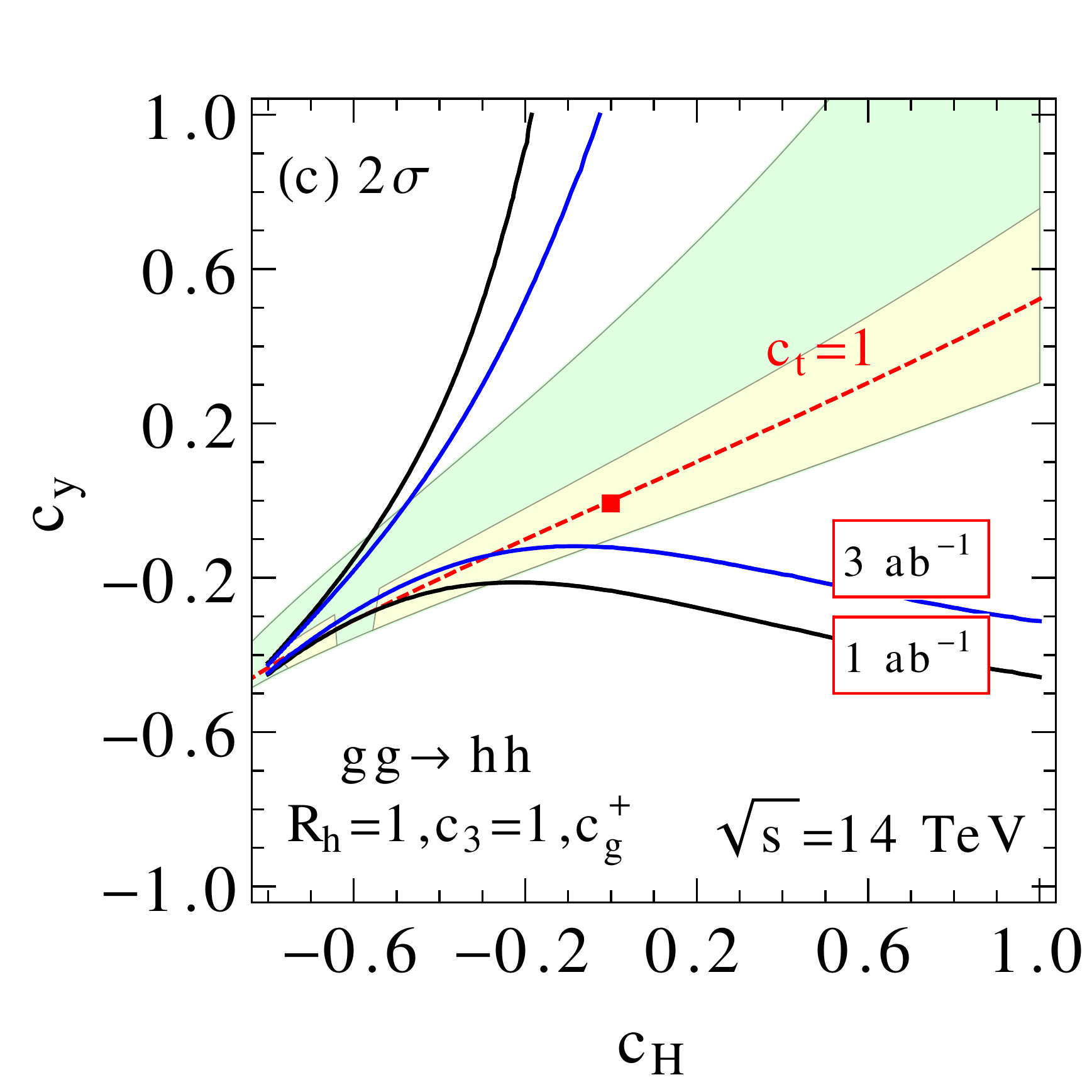}  
\includegraphics[scale=0.24]{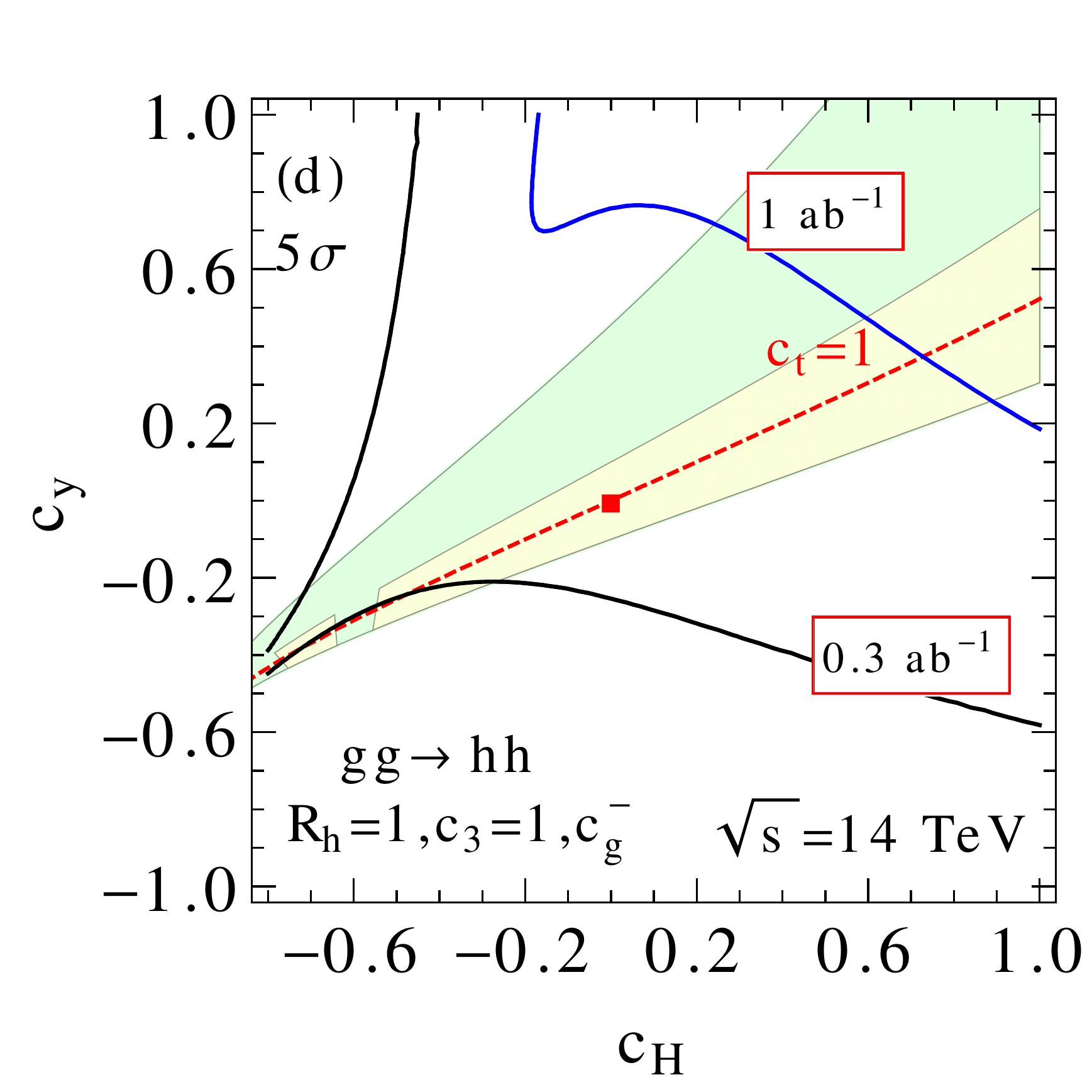}  
\caption{The $2 \sigma$ or $5\sigma$ discovery potential for $gg\to hh$ in the plane of $c_H$ and $c_y$ at the 14 TeV LHC. The green bands correspond to the constraints, at the $2\sigma$ C.L., from the measurements of $t\bar{t}h$ at the 13 TeV LHC.  
The gray and yellow bands represent the projected $2\sigma$ errors in the single Higgs production and  $t\bar{t}h$ measurement at the HL-LHC, respectively.
The red box denotes ($c_H=0$, $c_y=0$), while the red dashed line corresponds to $c_t=1$. $c_g^{\pm}$ refers to the sign choices "$\pm$" in Eq.~\eqref{eq:cg} with $R_h$ 
being fit from single Higgs production and decay data.
\label{fig:L14} }
\end{figure}

\begin{figure}
\includegraphics[scale=0.24]{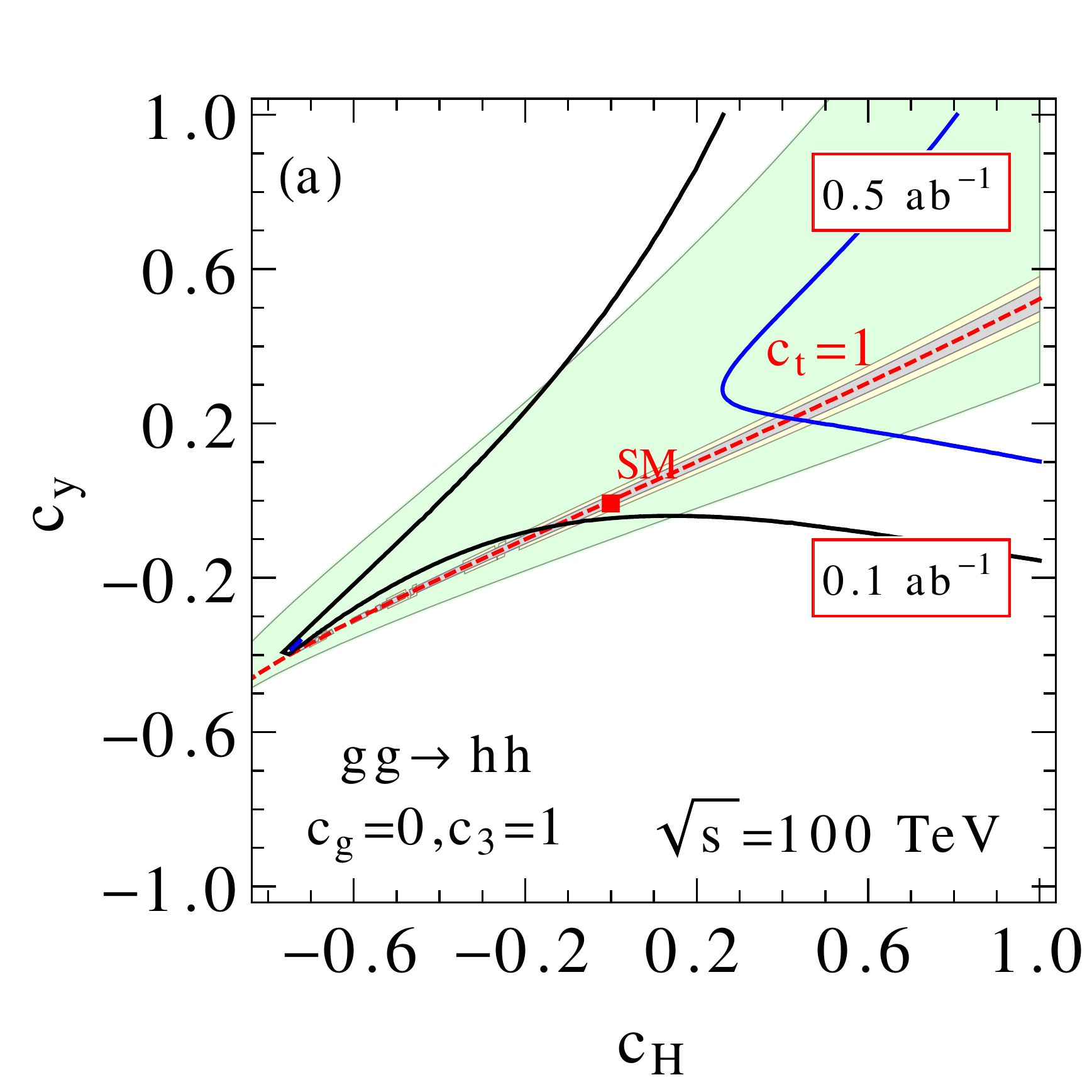}  
\includegraphics[scale=0.24]{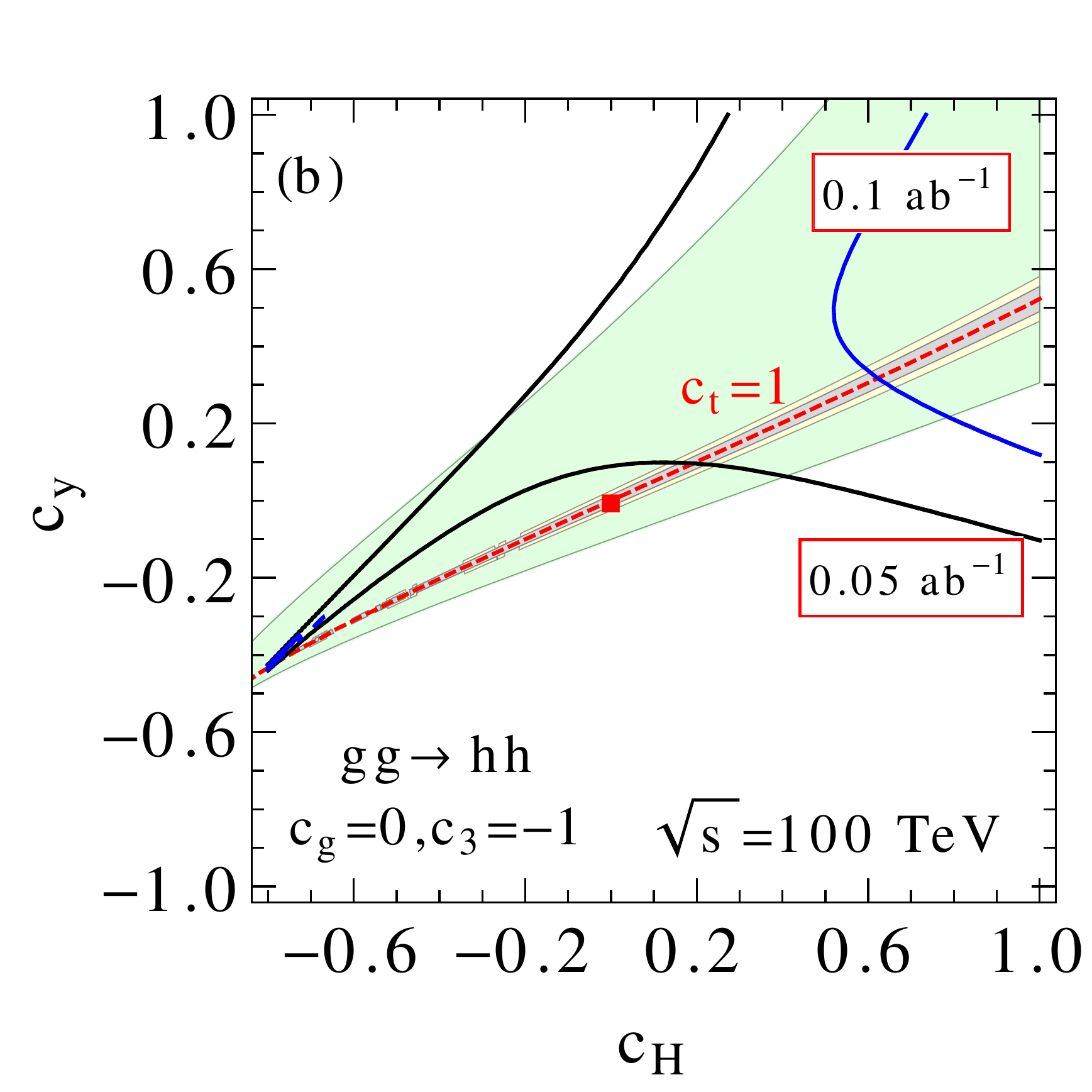}  
\includegraphics[scale=0.24]{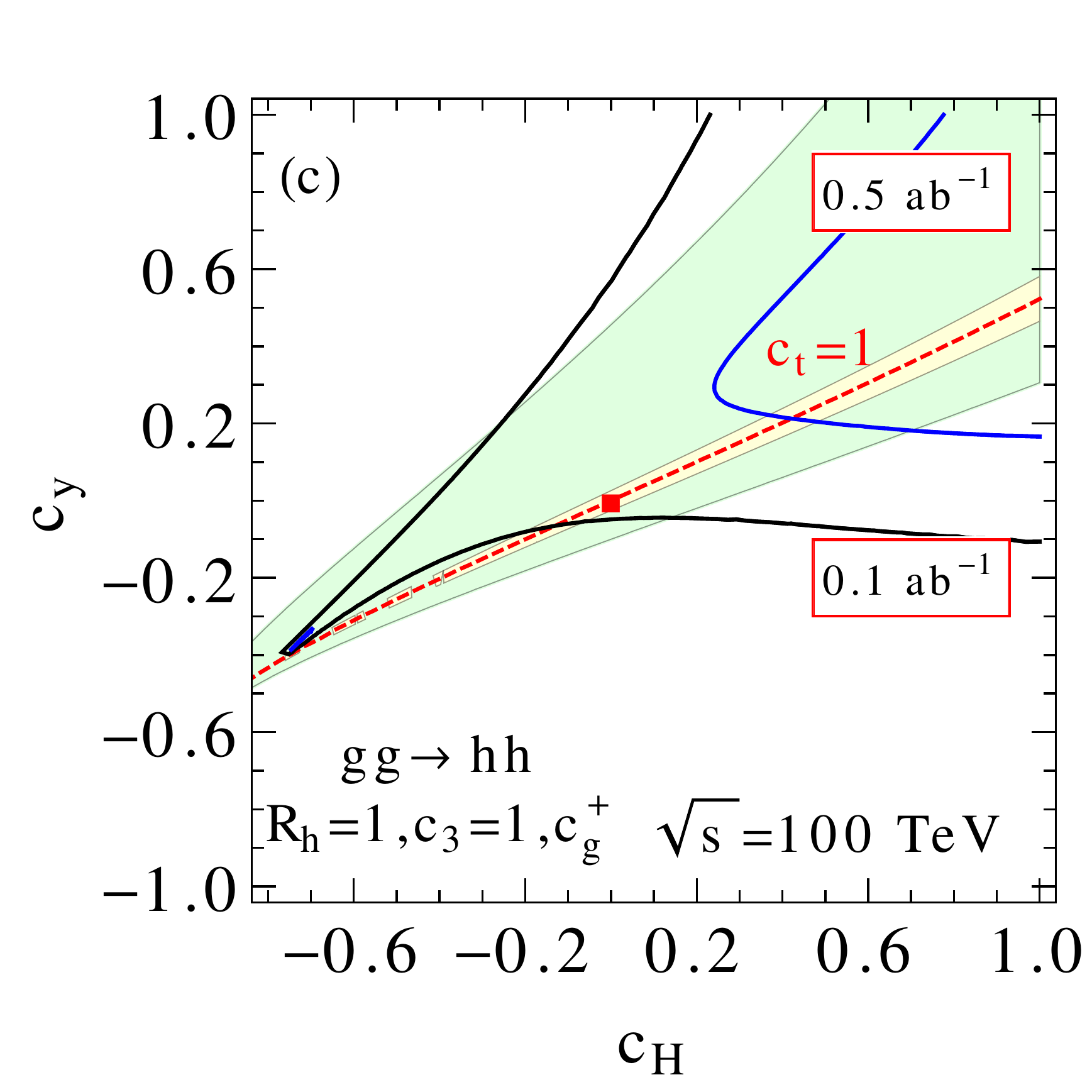}  
\includegraphics[scale=0.24]{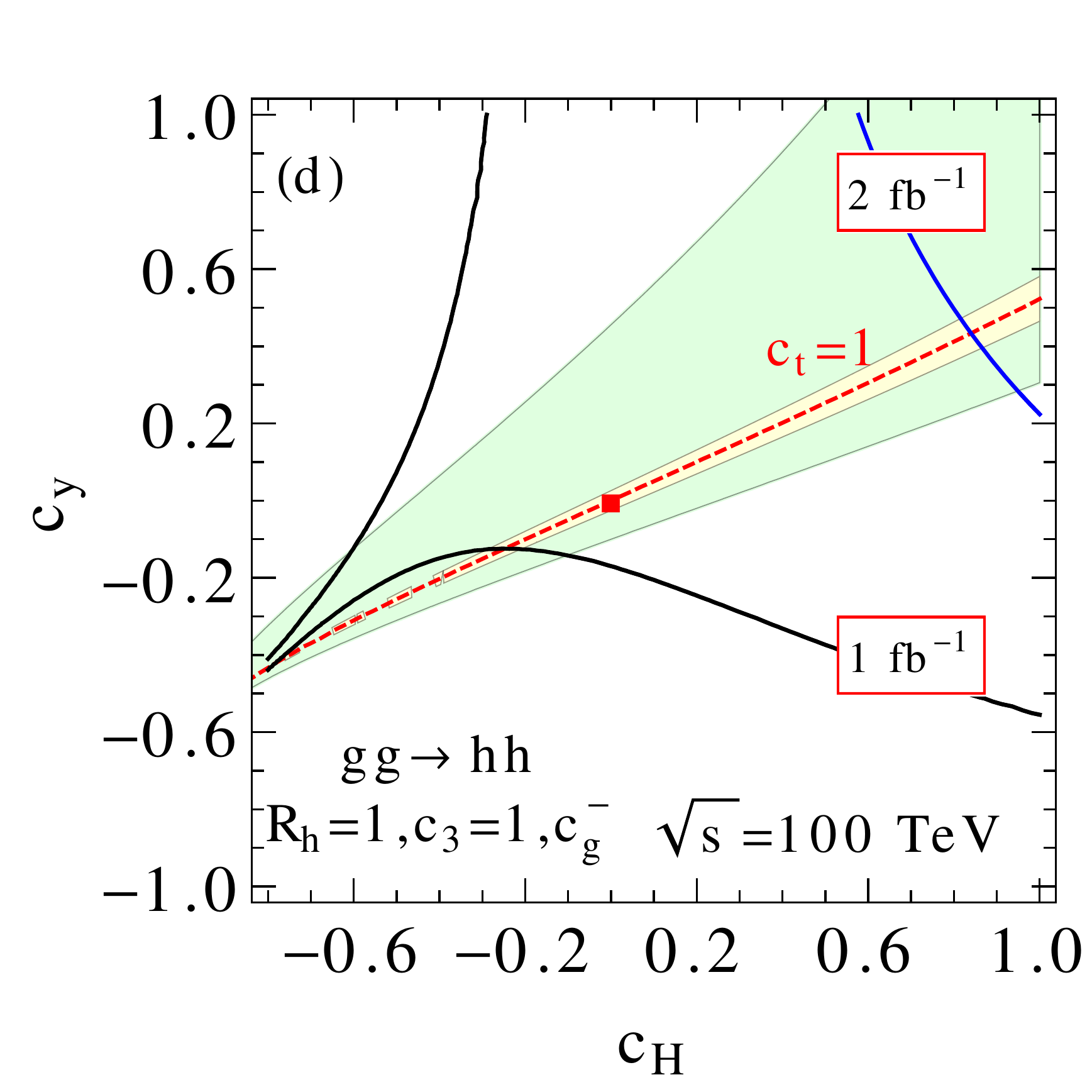}  
\caption{The $5 \sigma$ discovery potential for $gg\to hh$ in the plane of $c_H$ and $c_y$ at the 100 TeV hadron collider.  The green  bands correspond to the constraints, at the $2\sigma$ C.L., from the measurements of $t\bar{t}h$ at the 13 TeV LHC.  
The gray and yellow  bands represent the projected $2\sigma$ errors in the $t\bar{t}h$ measurement at the 100 TeV hadron collider, with an integrated luminosity of $3~{\rm ab}^{-1}$ at $2\sigma$ C.L., respectively.
The red box denotes ($c_H=0$, $c_y=0$), while the red dashed line corresponds to $c_t=1$. $c_g^{\pm}$ refers to the sign choices "$\pm$" in Eq.~\eqref{eq:cg} with $R_h$ being fit from single Higgs production and decay data.	
\label{fig:L100} }
\end{figure}

The SM backgrounds for the process of $gg\to hh$ production include $b\bar{b}\gamma\gamma$, $c\bar{c}\gamma\gamma$, $b\bar{b}\gamma j$, $jj\gamma\gamma$, $b\bar{b}jj$, $t\bar{t}(\geqslant 1\ell^\pm)$, $t\bar{t}\gamma$, $Z(\to b\bar{b})h(\to \gamma\gamma)$, $t\bar{t}h(\to \gamma\gamma)$ and $b\bar{b}h(\to \gamma\gamma)$, etc. At the 14~TeV LHC, with the integrated luminosity of $\mathcal{L}=3~{\rm ab}^{-1}$~\cite{ATL-PHYS-PUB-2014-019}, and the 100 TeV hadron collider, with $\mathcal{L}=30~{\rm ab}^{-1}$~\cite{Contino:2016spe}, the signal ($n_s$) and background ($n_b$) events in the SM are, respectively,
\begin{align}
14~{\rm TeV}&:  n_s=8.4, & &n_b=47,\nn\\
100~{\rm TeV}&:  n_s=12061, & & n_b=27118.
\end{align}
With the event numbers listed above, the discovery potential for the signal process can be evaluated by using~\cite{Cowan:2010js}
\begin{align}
\label{eq:discovery}
\mathcal{Z}&=\sqrt{2\bigg [(n_s+n_b)\log\dfrac{n_s+n_b}{n_b}-n_s\bigg ]}.
\end{align}
In Figs.~\ref{fig:L14} and~\ref{fig:L100}, we show the contours of  discovery potential for $gg\to hh\to b\bar{b}\gamma\gamma$ at the 14 TeV LHC and the 100 TeV hadron collider, with various integrated luminosities. The $2 \sigma$ and $5 \sigma$ discovery potentials correspond to $\mathcal{Z}=2$ and $\mathcal{Z}=5$, respectively.
As mentioned earlier, the signal strength $R_h$ is measured with an accuracy of about $10\%$, and $R_{hh}$ is not sensitive to the value of $R_h$ (for $0.9\leq R_h\leq 1.1$), we therefore take $R_h=1$ 
as the benchmark to show the discovery potential in those figures.

Several comments are in order regarding the discovery potential of $gg\to hh$ with respect to the Wilson coefficients $c_H$ and $c_y$. 
In Fig.~\ref{fig:L14}, only the parameter space on the left of the curve, or below the curve, labeled by a specified integrated luminosity at the HL-LHC, 
can be reached at the $2\sigma$ or $5\sigma$ C.L..
For case (a), the SM point cannot be probed by  measuring only the $gg\to hh$ production. 
A larger parameter space is reachable at $2\sigma$ C.L. for case $(b)$, as compared to cases (a) and (c), due to the large enhancement of $R_{hh}$ with negative $c_3$. 
For case $(d)$, with negative $c_g$, the Higgs pair production cross section can be enhanced by a factor of about 10, as compared to the SM value (cf. Fig.~\ref{fig:R14}$(f)$). Consequently, 
with an integrated luminosity of $1~{\rm ab}^{-1}$ at the HL-LHC, 
most of the considered parameter space (for $|c_H|<1$ and $|c_y|<1$) can already be reached at the 
$5\sigma$ C.L..
For comparison, in the same figure, we also show 
the constraints imposed by the projected $2\sigma$ errors in the single Higgs production and $t\bar{t}h$ measurement at the HL-LHC, with an integrated luminosity of $3~{\rm ab}^{-1}$.
It amounts to $c_t=1\pm 0.02$ (single Higgs with $c_g=0$) and 
$c_t=1\pm 0.05$ ($t\bar{t}h$ production)~\cite{ATL-PHYS-PUB-2014-016} which have included  the statistical and experimental systematic uncertainties.

In Fig.~\ref{fig:L100}, we see that the discovery potential for the $gg\to hh$ measurement is much improved at the 100 TeV hadron collider, for all the benchmark cases. The $5\sigma$ discovery significance can be easily reached. 
For cases $(a)$ and $(c)$, 
large region of $c_y,c_H< 0.2$ can be discovered with the integrated luminosity of $0.5~\text{ab}^{-1}$. For case $(b)$, with negative $c_3$, the region of $c_y,c_H< 0.5$ can be almost discovered with only $0.1~\text{ab}^{-1}$. For case $(d)$, with negative $c_g$, the currently allowed region can be explored with $2~\text{fb}^{-1}$. 
For comparison, in the same figure, we also show 
the constraints imposed by the projected $2\sigma$ errors in the single Higgs production and  $t\bar{t}h$ measurement at the 100 TeV hadron collider, with an integrated luminosity of $3~{\rm ab}^{-1}$.
It amounts to $c_t=1\pm 0.0072$ (single Higgs production, with $c_g=0$) and
$c_t=1\pm 0.013$ ($t\bar{t}h$ production)~\cite{Plehn:2015cta,Mangano:2651294}.  
Note that both the  statistical and systematic uncertainties are included in the single Higgs analysis, while only statistical error is discussed in $t\bar{t}h$ production.
Here, we have scaled down the error in $t\bar{t}h$ measurement by the inverse of the square root of integrated luminosity.

\begin{figure}
\includegraphics[scale=0.24]{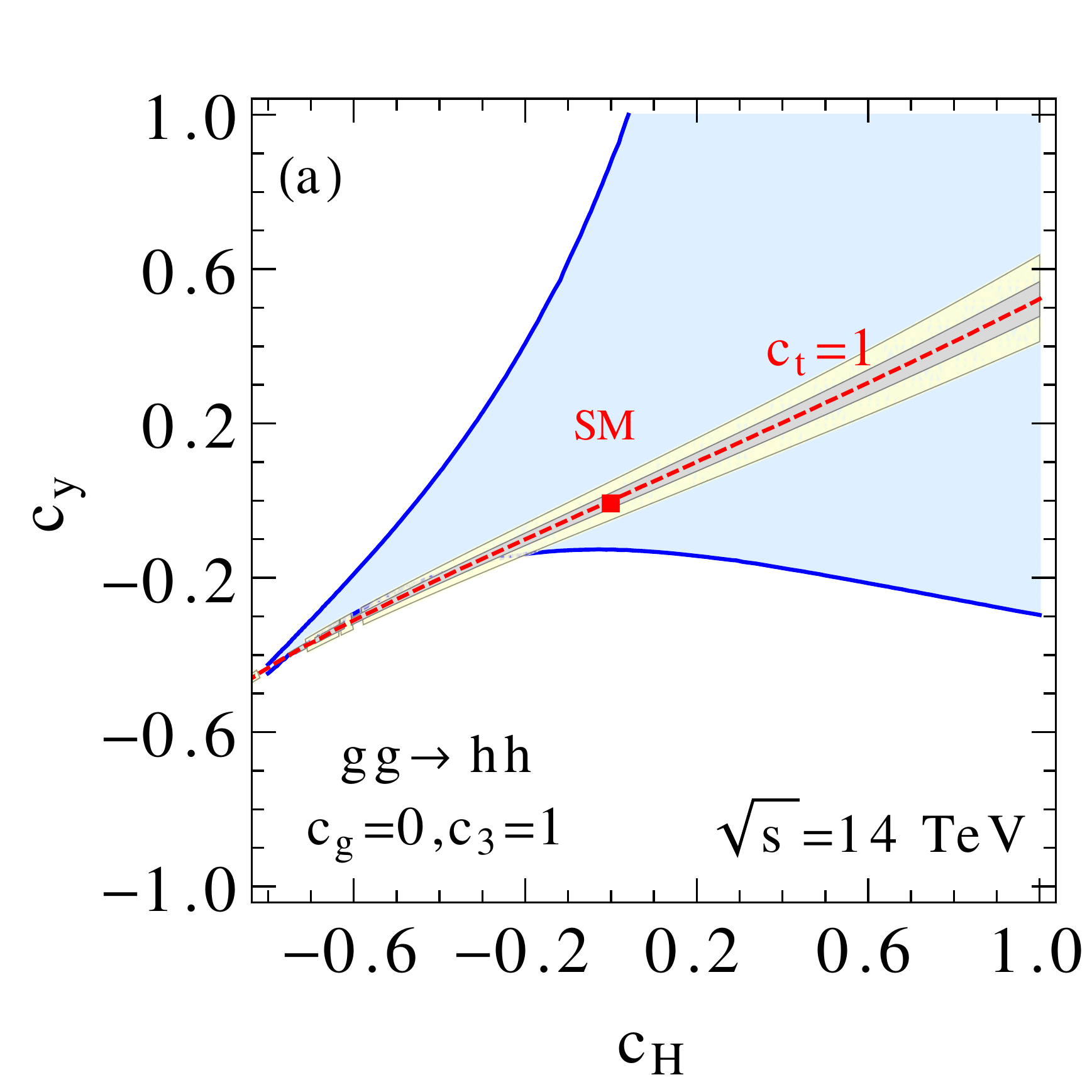}    
\includegraphics[scale=0.24]{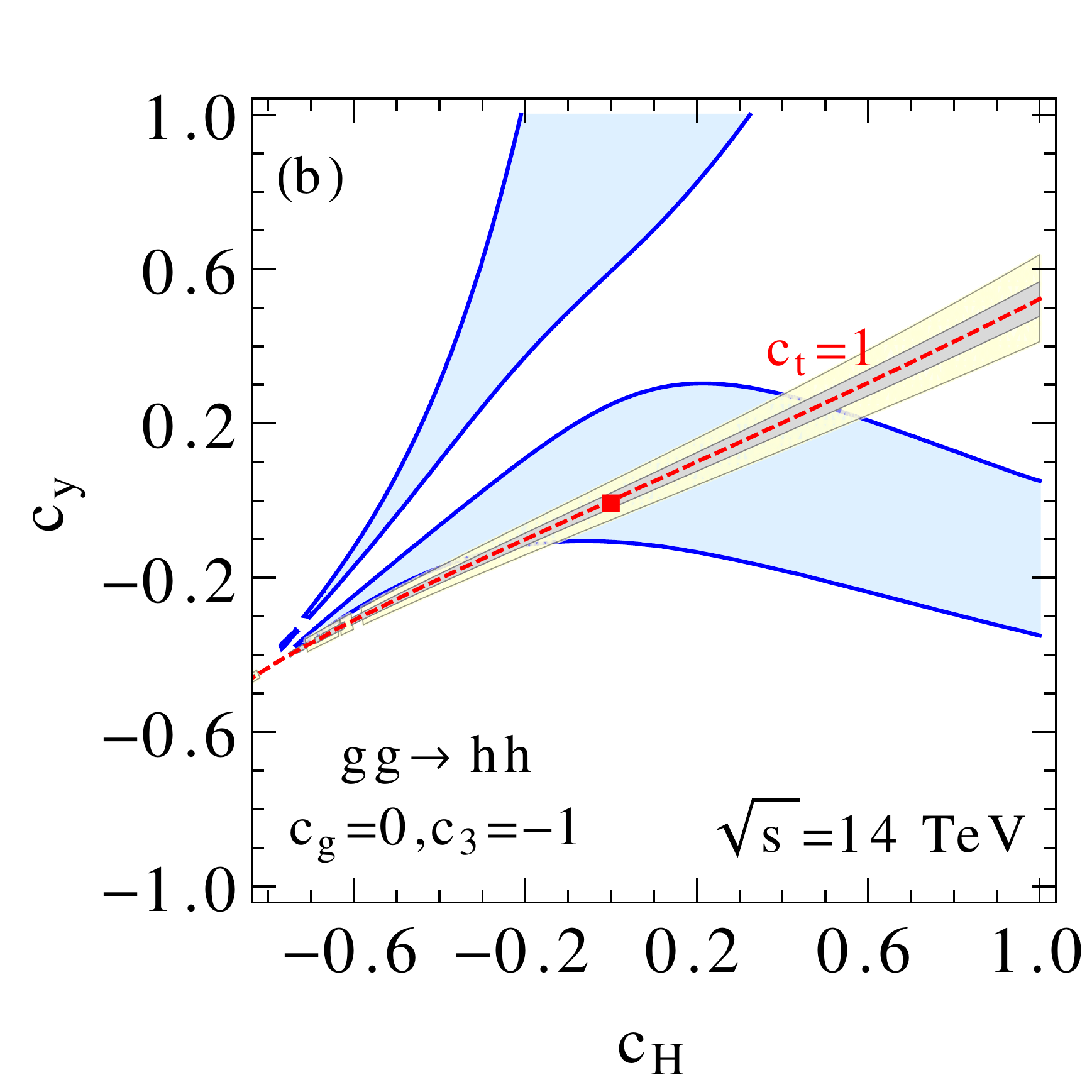}  
\includegraphics[scale=0.24]{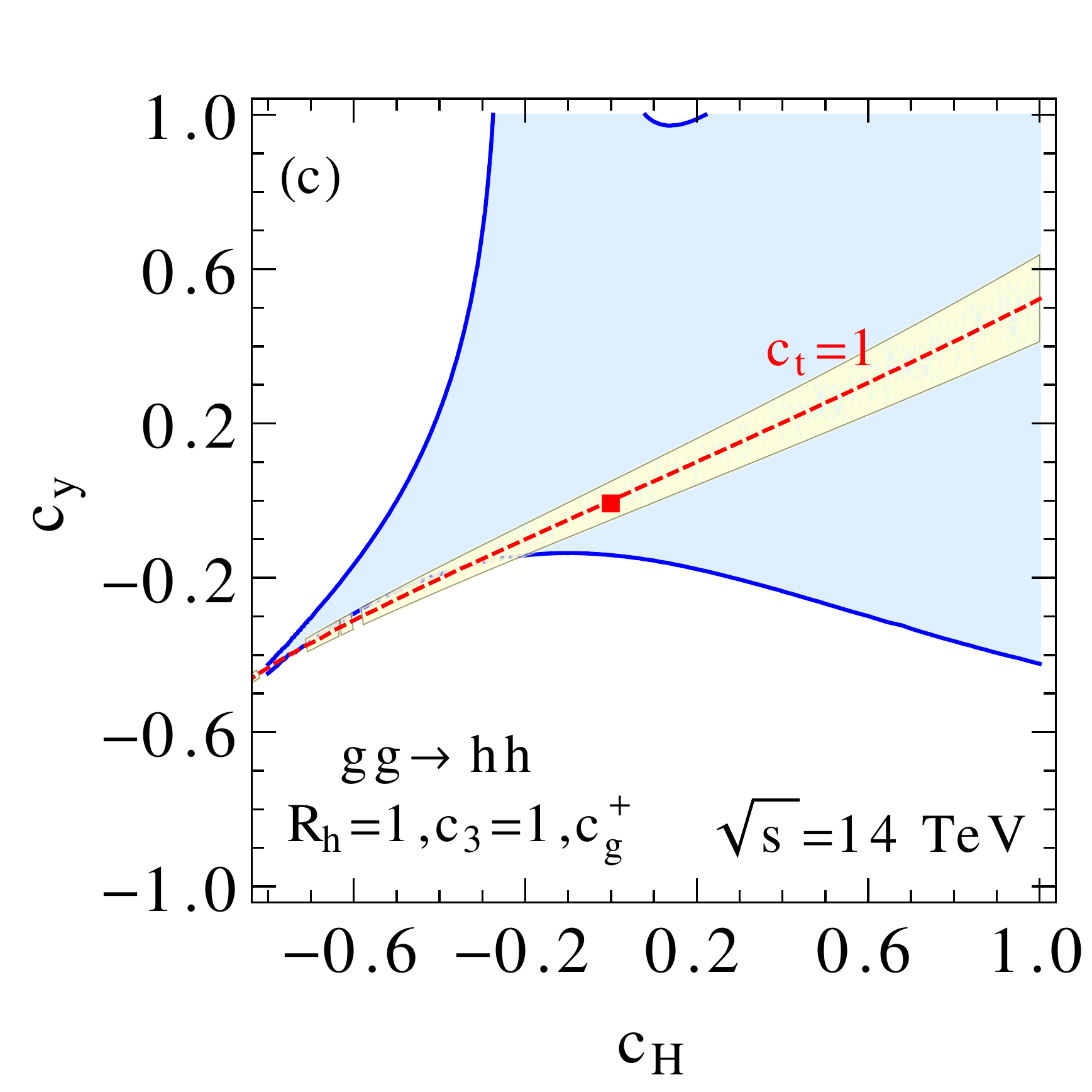}  
\includegraphics[scale=0.24]{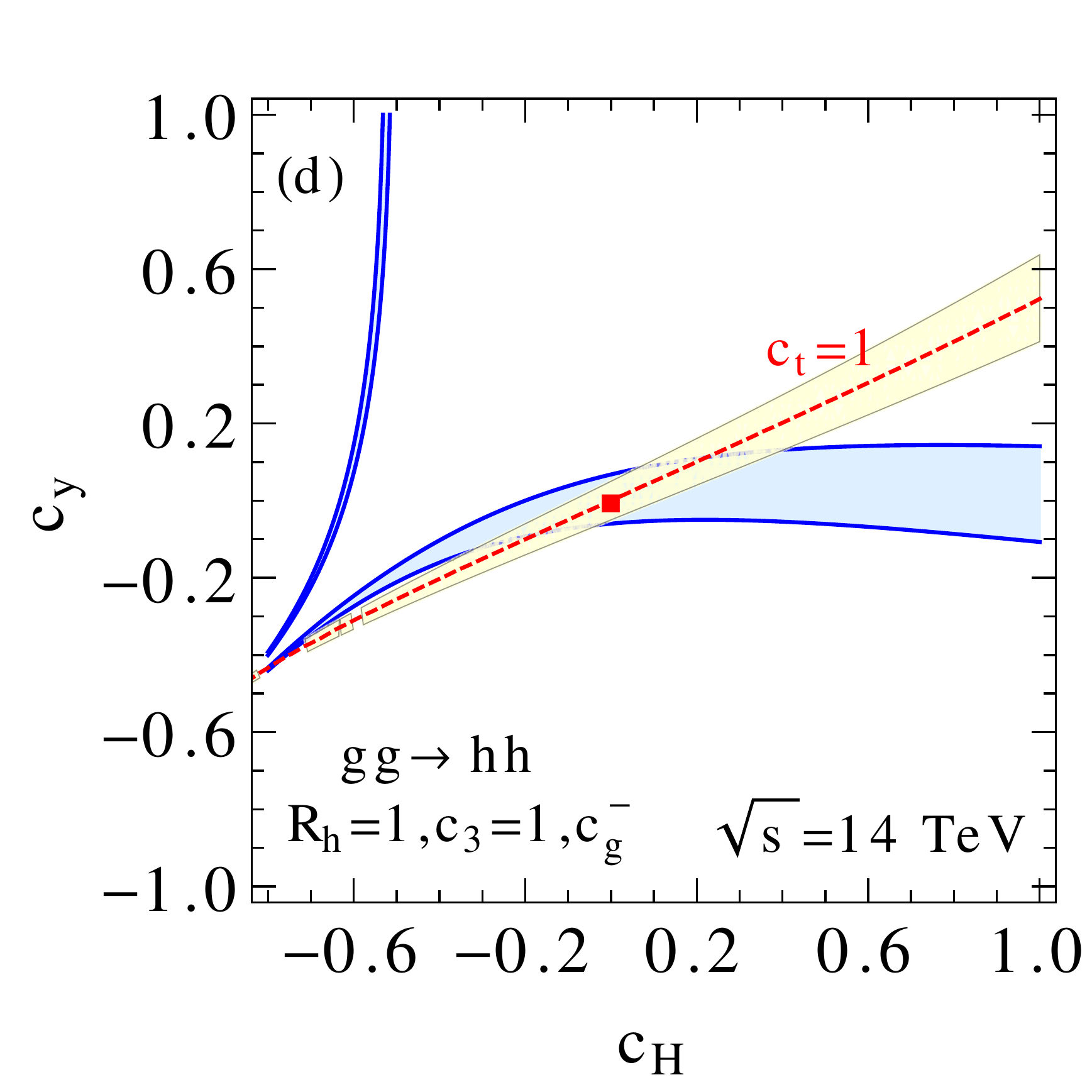}  
\caption{Expected accuracy for measuring $(c_H,c_y)$ with single Higgs production, $t\bar{t}h$ production and the Higgs pair production at the HL-LHC, with the integrated luminosity of $3~{\rm ab}^{-1}$. The gray, yellow  and blue bands represent the $1\sigma$ constraint from the measurements of single Higgs production,  $t\bar{t}h$ production and Higgs pair production  at the HL-LHC, respectively. Both the statistical and experimental systematic uncertainties have been included in the single Higgs production and $t\bar{t}h$ production. The red box denotes  ($c_H=0$, $c_y=0$), while the red dashed line coresponds to $c_t=1$.
\label{fig:precision14} }
\end{figure}

\begin{figure}
\includegraphics[scale=0.24]{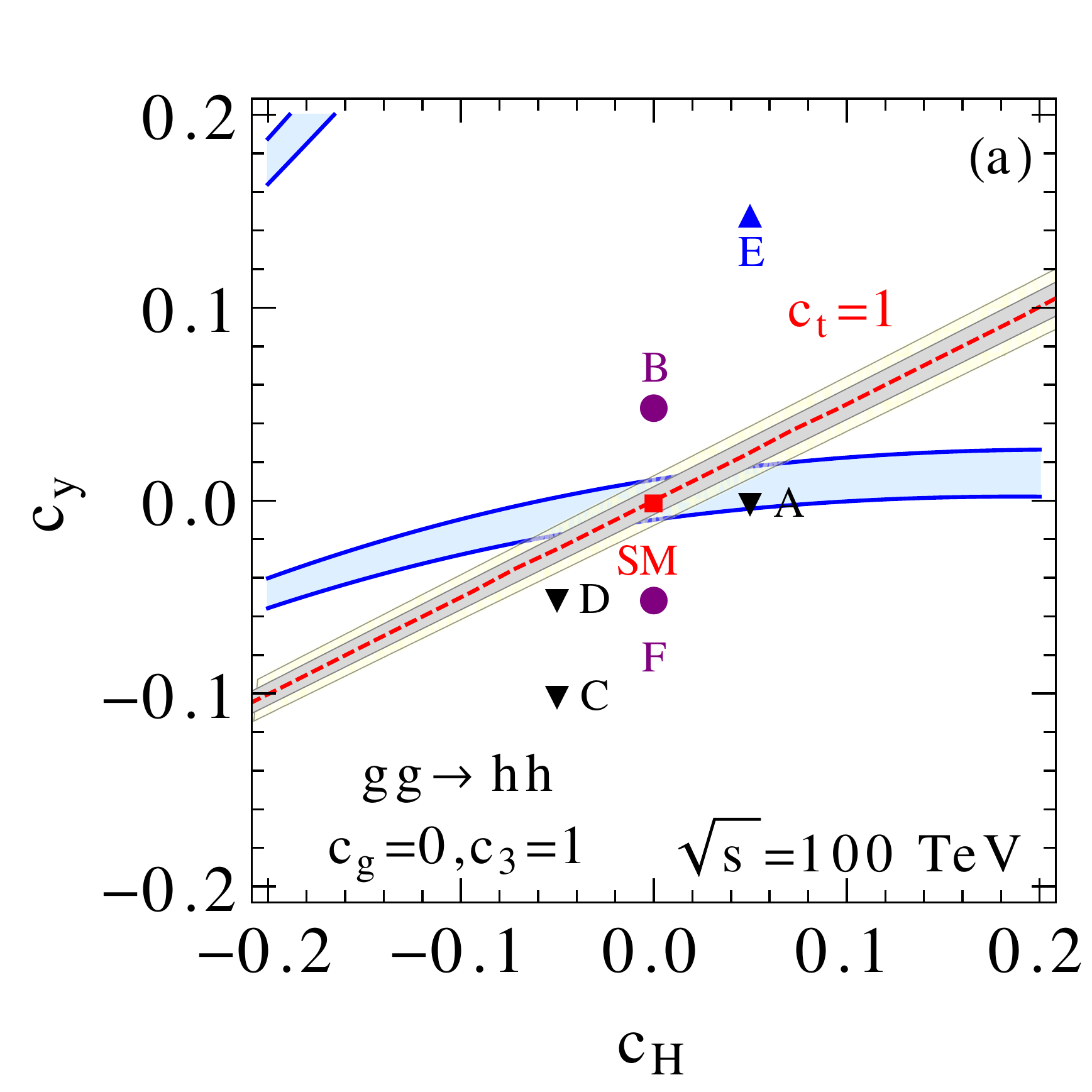}    
\includegraphics[scale=0.24]{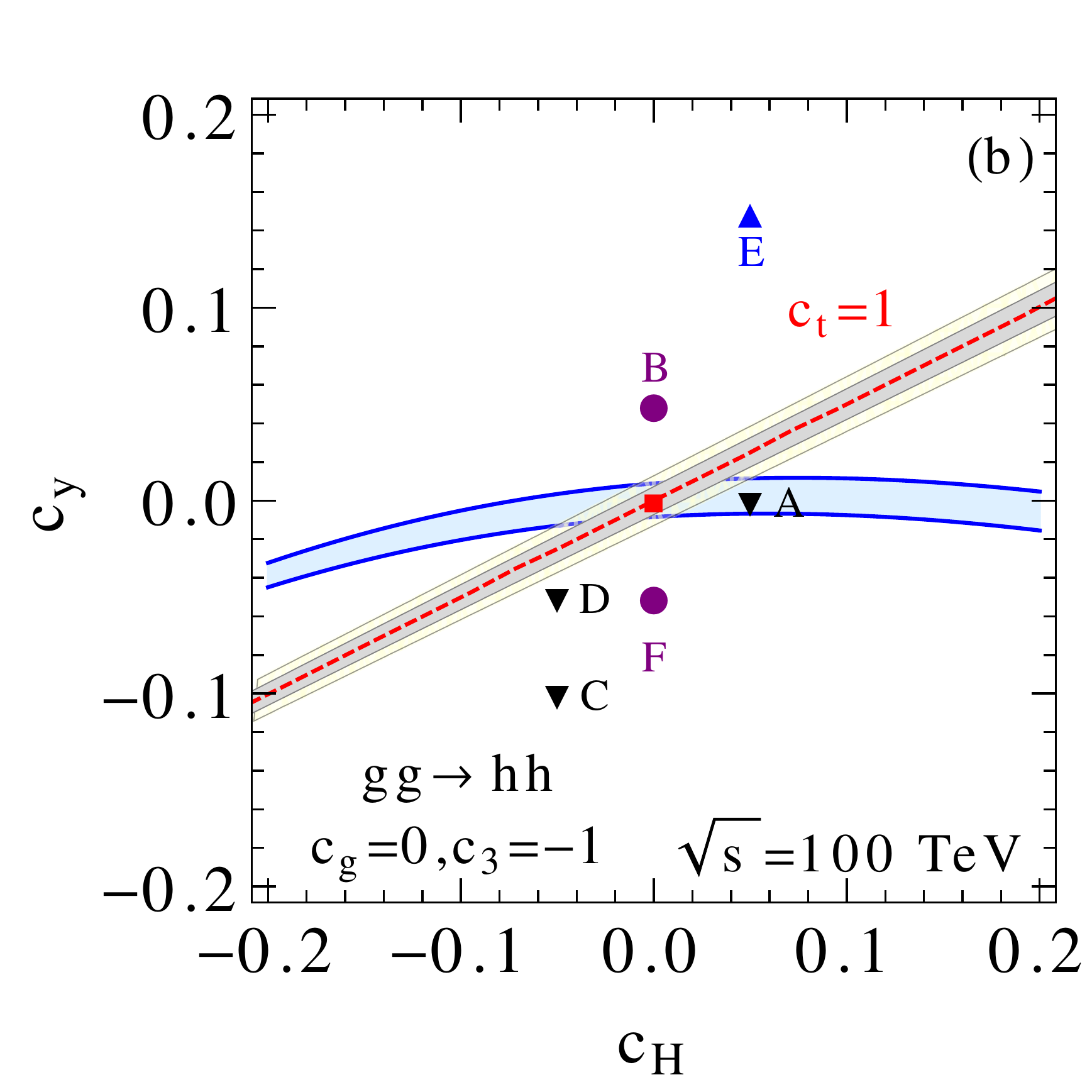}  
\includegraphics[scale=0.24]{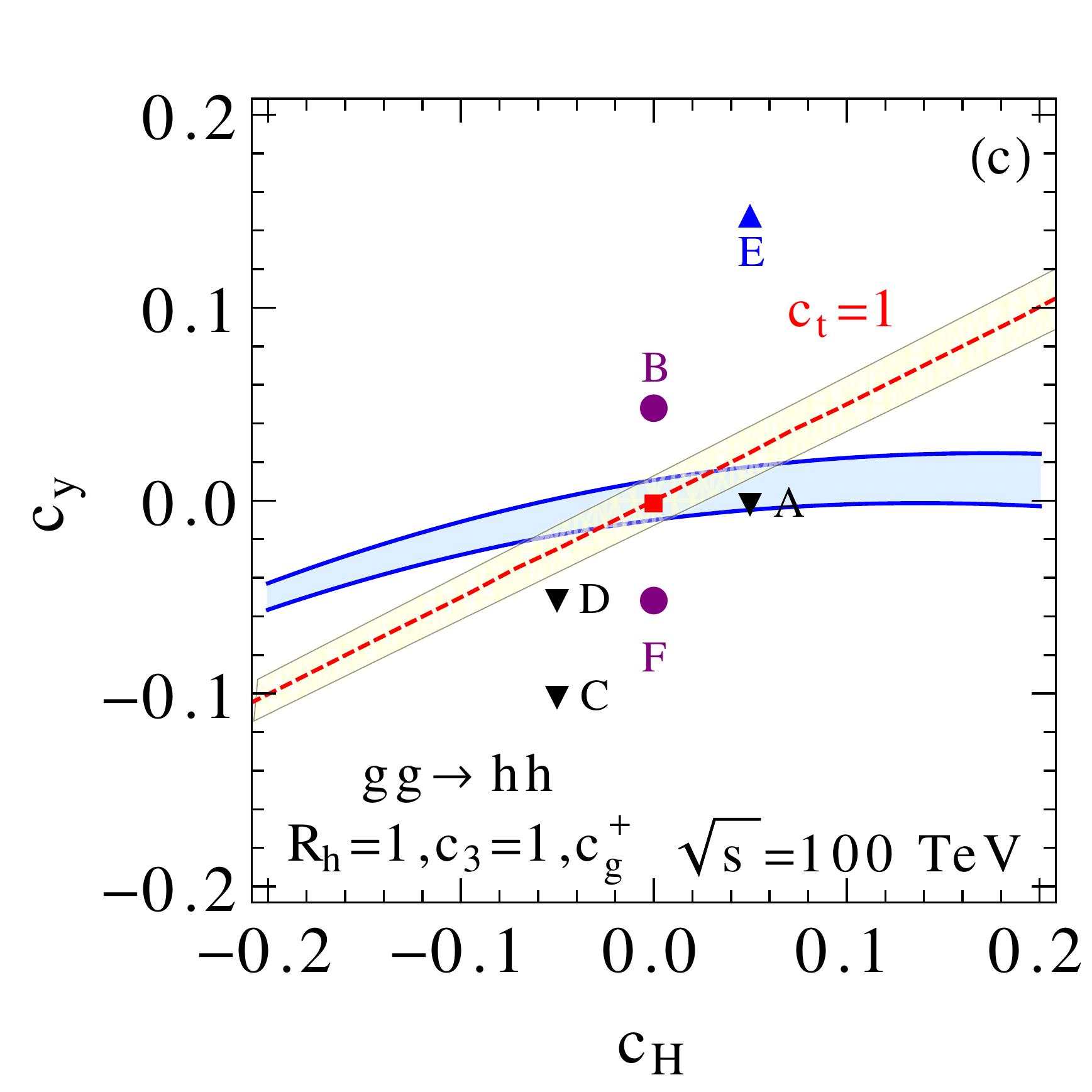}  
\includegraphics[scale=0.24]{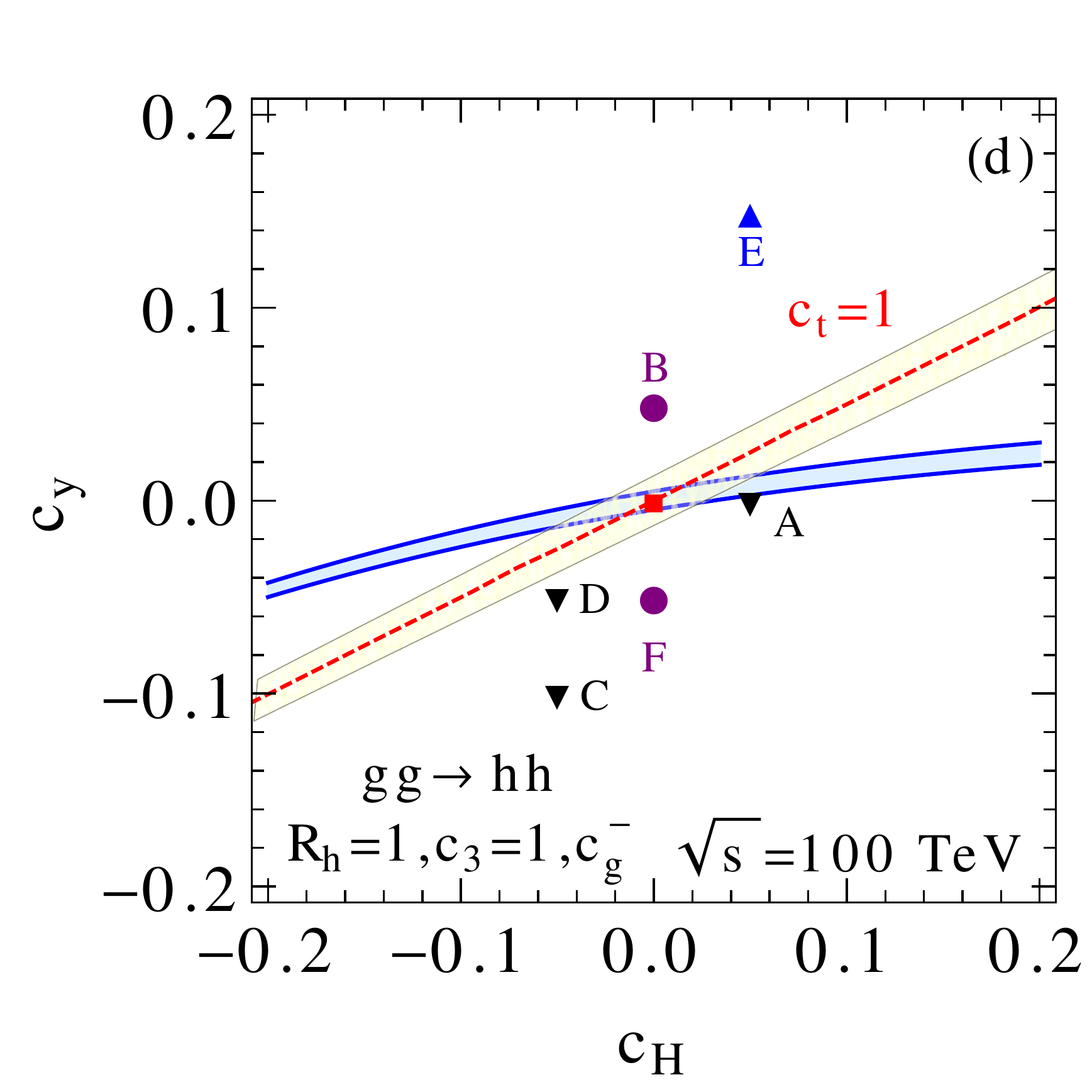}  
\caption{
Expected accuracy for measuring $(c_H,c_y)$ with single Higgs production, $t\bar{t}h$ production and the Higgs pair production at the 100 TeV collider, with the integrated luminosity of $3~{\rm ab}^{-1}$. The gray,  yellow and blue bands represent the $1\sigma$ constraint from the measurements of single Higgs production, $t\bar{t}h$ production and  Higgs pair production  at the 100 TeV collider, respectively. Both the statistical and experimental systematic uncertainties have been included in the single Higgs production. The red box denotes  ($c_H=0$, $c_y=0$), while the red dashed line coresponds to $c_t=1$.	
For comparison, several benchmark points of NP models are also shown: A) singlet scalar, B) 2HDM or VLQs with $c_y>0$, C) real triplet scalar, D) complex triplet scalar, E) vectors, F) 2HDM or VLQs with $c_y<0$.
\label{fig:precision} }
\end{figure}

To estimate the expected accuracy for measuring $(c_H,c_y)$  with the Higgs pair production at the 14~TeV LHC and the 100~TeV hadron collider, respectively, we perform a log likelihood ratio test~\cite{Cowan:2010js} for the hypothesis with non-zero $c_H,c_y$ against the hypothesis with $c_H=c_y=0$. The test ratio is defined as~\cite{Cowan:2010js}
\begin{align}
t=-2\ln\dfrac{L(c_H,c_y)}{L(0,0)}
\end{align}
where the likelihood function $L(c_H,c_y)$ is 
\begin{align}
L(c_H,c_y)=P(\text{data}|n_b+n_s(c_H,c_y)).
\end{align}
Here $P(k|\lambda)$ is the usual Poisson distribution function, $P(k|\lambda)=\lambda^k e^{-\lambda}/k!$. We assume the observed data is generated under the hypothesis with $c_H=c_y=0$~\cite{Cowan:2010js,Kumar:2015tna} and calculate the two-sided $p$-value.
For convenience, we convert the $p$-value into the equivalent significance $Z=\Phi^{-1}(1-1/2p)=\sqrt{2}~\text{Erf}^{-1}(1-p)$~\cite{Tanabashi:2018oca,Cowan:2010js}, where $\Phi$ the cumulative distribution of the standard Gaussian and $\text{Erf}$ is the error function. In our case, 
$Z=\sqrt{2\big[n_0\ln\dfrac{n_0}{n_1}+(n_1-n_0)\big]}$ with $n_0=n_b+n_s(0,0)$ and $n_1=n_b+n_s(c_H,c_y)$.
The discrimination between the hypothesis with arbitrary $(c_H,c_y)$ and the hypothesis with $c_H=c_y=0$ is shown in Figs.~\ref{fig:precision14} and~\ref{fig:precision}. The hypothesis with $(c_H,c_y)$ outside of the blue bands is rejected at $1\sigma$ 
level for the HL-LHC and the 100~TeV hadron collider, respectively. 
After combining the measurements of single Higgs production (gray band) and $t\bar{t}h$ production (yellow band) at the HL-LHC, it is possible to differentiate $c_H$ and $c_y$ at the HL-LHC  for cases $(b)$ and $(d)$, but it becomes challenging for cases $(a)$ and $(c)$, cf. Fig.~\ref{fig:precision14}. 
The situation will be much improved  at the 100 TeV hadron collider, cf. Fig.~\ref{fig:precision}.
It is obvious that the Higgs pair production is more sensitive to $c_y$ than to $c_H$.
We find that 
at the $1\sigma$ C.L.,  
the combined constraint from single Higgs production, $t\bar{t}h$ production and the Higgs pair production measurements, at the 100~TeV hadron collider with the integrated luminosity of $3~\text{ab}^{-1}$, yields 
$-0.04< c_y< 0$ and $-0.08< c_H< 0.08$. 
To further constrain  $c_H$, it is necessary to improve the measurement of $t\bar{t}h$ associated production.
However, if the Higgs boson is a SM-like particle, $c_H$ could be constrained by the $hVV$ coupling measurement to $1\%\sim2\%$ level~\cite{Mangano:2651294}.

Given the good sensitivity of differentiating $c_H$ with $c_y$ at the 100~TeV hadron collider, it is worthwhile clarifying the specific values of $(c_H,c_y)$, as induced by several generic classes of NP  models~\cite{Corbett:2017ieo,delAguila:2000rc,Low:2009di}.
Table~\ref{tbl:coupling} lists 
the Wilson coefficients $c_H$ and $c_y$ predicted by various NP models~\cite{delAguila:2000rc,Low:2009di,Corbett:2017ieo}. Both heavy scalars and vectors could contribute to $c_H$ and $c_y$, while the additional heavy vector-like quarks (VLQs) could contribute to $c_y$ and $c_g$. The sign of $c_y$ is arbitrary in two Higgs doublet (2HDM)~\cite{Corbett:2017ieo} and VLQ models~\cite{delAguila:2000rc}. 
Those NP models can be easily discriminated if they modify $c_H$ or $c_y$ by a sizable amount, cf. Fig.~\ref{fig:precision}.

\begin{table}
	\caption{The Wilson coefficients $c_H$ and $c_y$ predicted by various NP  models~\cite{delAguila:2000rc,Low:2009di,Corbett:2017ieo}.}
	\begin{center}
		\begin{tabular}{c|c}
			\hline\hline
			 A) singlet scalar~\cite{Low:2009di,Corbett:2017ieo} & B) 2HDM~\cite{Corbett:2017ieo} or VLQs~\cite{delAguila:2000rc}  \\
			\hline
			 $c_H>0, c_y=0 $ & $c_H=0, c_y>0$  \\
			 \hline\hline
			 C) real triplet scalar~\cite{Low:2009di}   &  D) complex triplet scalar~\cite{Low:2009di} \\
			\hline
			$c_H=2 c_y<0$ &  $c_H=c_y<0$ \\
			\hline\hline
		         E) vectors~\cite{Low:2009di} & F) 2HDM~\cite{Corbett:2017ieo} or VLQs ~\cite{delAguila:2000rc}  \\
			\hline
			 $c_H=3 c_y>0$ &$c_H=0, c_y<0$ \\
			\hline\hline
		\end{tabular}
	\end{center}
	\label{tbl:coupling}
\end{table}

\noindent{\bf Conclusions.}
Both the Wilson coefficients $c_H$ and $c_y$ of dimension-six operators can contribute to the top quark Yukawa coupling simultaneously, thus their individual contributions cannot be separated with the measurement of $ht\bar{t}$ coupling alone. 
In this work, we demonstrate that $c_H$ and $c_y$ also contribute to the $t\bar{t}hh$ effective coupling $c_{2t}$, whose information can be well extracted out from the Higgs pair production. Thus this process can be used to distinguish the effects of $c_H$ and $c_y$ at the 14 TeV LHC and the 100 TeV hadron collider. Regarding the discovery potential for the process $gg\to hh$, it shows that the $2\sigma$ confidence level is reachable for some parameter space at the 14 TeV LHC in general, and the sensitivity can be much improved at the 100 TeV hadron collider. Regrading the sensitivity of measuring $c_t$ and $c_H$, we find it is challenging to differentiate $c_H$ and $c_y$ at the HL-LHC, except for some special scenarios. The situation will be much improved at the 100 TeV hadron collider. After combing the single Higgs production, $t\bar{t}h$ production at the 100 TeV hadron collder, the Higgs pair production can give a strong constraint on $c_y$ regardless of the value $c_H$.  The precise measurement of both $c_H$ and $c_y$ enable us to discriminate various new physics models.

\vspace*{3mm}

\noindent{\bf Acknowledgements.}
We thank Kirtimaan A. Mohan for helpful discussions. The work of 
G. Li is supported by the MOST (Grant No. MOST 106-2112-M-002-003-MY3).  
L.-X. Xu is supported in part by the National Science Foundation of China under Grants No. 11635001, 11875072.
B. Yan and C.-P. Yuan  are supported by the U.S. National Science Foundation under Grant No. PHY-1719914. C.-P. Yuan is also grateful for the support from the Wu-Ki Tung endowed chair in particle physics.

\bibliographystyle{apsrev}
\bibliography{reference}

\end{document}